\documentclass[prb,superscriptaddress,notitlepage,aps,reprint]{revtex4-1}

\usepackage{graphicx,color,bm,amsmath,amssymb,natbib}
\newcommand{\UPt}{UPt$_3$}

\begin{document}
\title{Reversible ordering and disordering of the vortex lattice in {\UPt}}

\author{K.~E.~Avers}
\altaffiliation[Current address: ]{Department of Physics, University of Maryland, College Park, Maryland 20742, USA}
\affiliation{Department of Physics and Astronomy, Northwestern University, Evanston, Illinois 60208, USA}
\affiliation{Center for Applied Physics \& Superconducting Technologies, Northwestern University, Evanston, Illinois 60208, USA}

\author{S.~J.~Kuhn}
\altaffiliation[Current address: ]{Center for Exploration of Energy \& Matter, Indiana University, Bloomington, Indiana 47408, USA}
\affiliation{Department of Physics and Astronomy, University of Notre Dame, Notre Dame, Indiana 46556, USA}

\author{A.~W.~D.~Leishman}
\affiliation{Department of Physics and Astronomy, University of Notre Dame, Notre Dame, Indiana 46556, USA}

\author{W.~J.~Gannon}
\altaffiliation[Current address: ]{Department of Physics and Astronomy, University of Kentucky, Lexington, Kentucky 40506, USA}
\affiliation{Department of Physics and Astronomy, Northwestern University, Evanston, Illinois 60208, USA}

\author{L.~DeBeer-Schmitt}
\affiliation{Large Scale Structures Section, Neutron Scattering Division, Oak Ridge National Laboratory, Oak Ridge, Tennessee 37831, USA}

\author{C.~D.~Dewhurst}
\author{D.~Honecker}
\author{R.~Cubitt}
\affiliation{Institut Laue-Langevin, 71 avenue des Martyrs, CS 20156, F-38042 Grenoble cedex 9, France}

\author{W.~P.~Halperin}
\affiliation{Department of Physics and Astronomy, Northwestern University, Evanston, Illinois 60208, USA}

\author{M.~R.~Eskildsen}
\email[email: ]{eskildsen@nd.edu}
\affiliation{Department of Physics and Astronomy, University of Notre Dame, Notre Dame, Indiana 46556, USA}

\date{\today}

\begin{abstract}
When studied by small-angle neutron scattering the vortex lattice (VL) in {\UPt} undergoes a gradual disordering as a function of time due to $^{235}$U fission.
This temporarily heats regions of the sample above the critical temperature, where, upon re-cooling, the vortices remain in a quenched disordered state.
The disordering rate is proportional to the magnetic field, suggesting that it is governed by collective VL properties such as the elastic moduli.
An ordered VL can be re-formed by applying a small field oscillation, showing that the fission does not cause detectable radiation damage to the {\UPt} crystals, even after long exposure.
\end{abstract}

\date{\today}

\maketitle

\section{Introduction}
Quantized vortices are introduced in a type-II superconductor subjected to an applied magnetic field~\cite{Huebener:2001we}.
Understanding and controlling vortex matter is of both fundamental interest and practical importance since vortex motion leads to dissipation.
In an idealized scenario, vortices will arrange themselves in a perfectly ordered vortex lattice (VL) due to their mutual repulsion~\cite{Abrikosov:1957vu,Kleiner:1964ih,Matricon:1964bt}.
In reality, however, thermal effects and/or pinning to material defects is always present, and the balance between these competing factors determines the structural and dynamic properties of vortex matter~\cite{Blatter:1994gz,Brandt:1995aa,Giamarchi:1995tq,Giamarchi:1997th,Nattermann:2000vk,LeDoussal:2010wu}.
This leads to a complex, high-dimensional phase diagram, where transitions between different states are driven not only by changes in intensive quantities, such as the field or temperature, but also the amount of imperfection or impurities which affect the vortex pinning.
An example of the latter is columnar defects introduced by heavy ion irradiation~\cite{Menghini:2003fe}.
In many applications control of vortex dynamics is of critical importance such as for high coherence in superconducting RF cavities used in accelerators~\cite{Checchin:2020ko} and for quantum information science~\cite{Romanenko:2020jj}.

The ability to manipulate the vortices experimentally is essential to the study of vortex matter.
Frequently, the VL configuration will depend on the field and temperature history, either in the degree of ordering~\cite{Xiao:2004gk,MarzialiBermudez:2015uj,MarzialiBermudez:2017vl} or the orientation of the VL relative to the crystalline axes of the host material~\cite{Huxley:2000aa,Das:2012cf,Okuma:2012dz}.
In materials with weak vortex pinning, it is possible to anneal quenched disorder or dislodge the system from an ordered but metastable configuration by temporarily exciting the VL, either by applying a transport current~\cite{Yaron:1994tb,Yaron:1995wa,Duarte:1996gl,Pautrat:2005ku,Li:2006cg} or a small-amplitude magnetic field oscillation~\cite{Levett:2002ba,Louden:2019bq,Louden:2019io}.
This causes vortex motion and ``shakes'' them free of local minima in their collective energy landscape.
In contrast, vortex matter in superconductors with strong pinning often become more disordered by the application of a vortex shaking~\cite{Eskildsen:2009cx,Inosov:2010aa,Inosov:2010bb}.

Here we demonstrate a novel approach to structural studies of vortex matter whereby reversible quenched disorder can be introduced locally without permanently affecting the host superconducting material.
Specifically, we used small-angle neutron scattering (SANS) to study the VL in the topological superconductor {\UPt}, which undergoes a gradual disordering on a time scale of tens of minutes as it is subjected to a beam of cold neutrons.
The disordering is due to local heating events caused by neutron induced fission of $^{235}$U, which leaves an increasing fraction of the sample occupied by a disordered vortex configuration.
While the system does not spontaneously re-order once the local heating has been dissipated, it is possible to re-anneal the VL by the application of a damped field oscillation.

\section{Experimental details}
\begin{figure*}
	\includegraphics{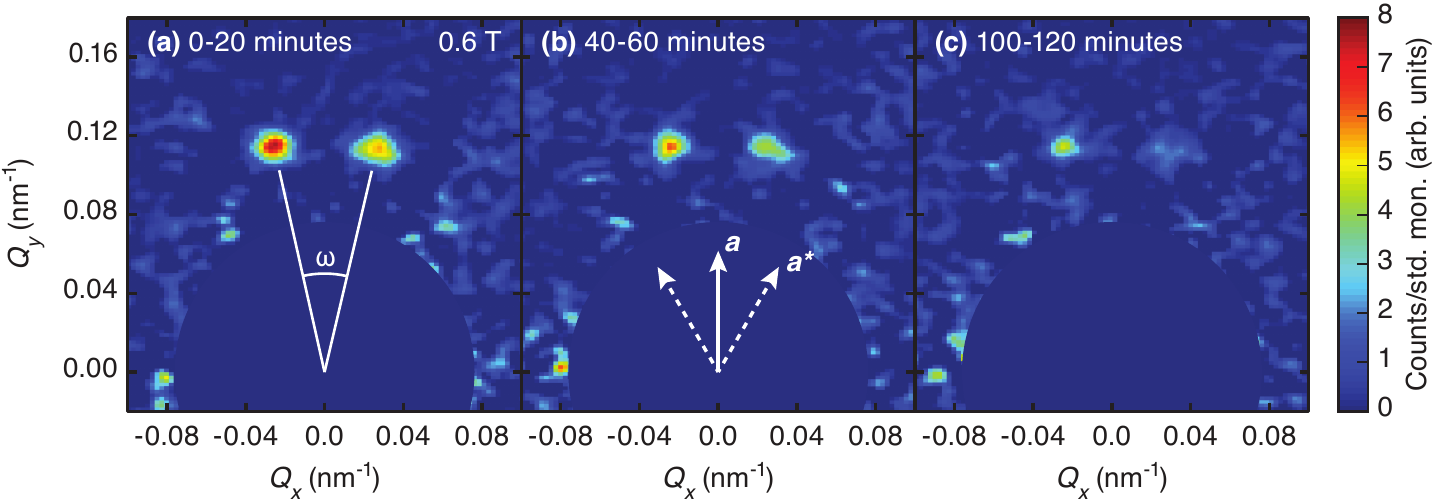}
	\caption{\label{DifPatDecay}
        Diffraction patterns obtained on ZR11 at 0.6~T (ORNL), measured at different times following the preparation of a pristine VL and without the application of periodic field oscillations.
        Each diffraction pattern was counted for 20 minutes and uses the same color scale.
        The peak splitting is indicated in (a) and crystallographic directions within the scattering plane in (b).
        Only peaks at the top of the detector satisfied the Bragg condition.
        Background scattering is subtracted, and the detector center near $Q = 0$ is masked off.
       }
\end{figure*}

The SANS measurements~\cite{Muhlbauer:2019jt} were performed on the CG-2 General Purpose SANS instrument at the High Flux Isotope Reactor at Oak Ridge National Laboratory (ORNL)~\cite{Heller:2018} and on the D33 instrument at the Institut Laue-Langevin (ILL)~\cite{Dewhurst:2014wo,5-42-402,5-42-467}, using a fixed neutron wavelength obtained by a velocity selector.
Two high-quality single crystals were studied, designated ZR8 and ZR11, both of which have been used in previous SANS measurements~\cite{Gannon:2015ct,Avers:2020wx,Avers:2022gh}.
The characteristics for each crystal is given in the Supplemental Material.
Unless otherwise stated, SANS measurements were carried out in a ``rocked on'' configuration, satisfying the Bragg condition for VL peaks at the top of the two-dimensional position sensitive detector, as seen in Fig.~\ref{DifPatDecay}.
Measurements were performed using a dilution refrigerator operating at base temperature of $50-65$~mK $\sim 0.1 T_c$
and with applied magnetic fields between 0.3~T and 1.0~T applied along the crystalline $\textbf{c}$ axis.
Background measurements, obtained either in zero field or above the upper critical field, were subtracted from the foreground data.

Prior to each SANS measurement sequence a pristine VL was prepared by applying a damped magnetic field oscillation with an initial amplitude of 20~mT around the measurement field.
This was previously found to produce a well ordered VL with a homogeneous vortex density~\cite{Avers:2020wx}.
In addition, a $\pm 5$~mT oscillation was applied periodically during some of the SANS measurements to maintain an ordered VL~\cite{Avers:2020wx}.
All field oscillations end with a reduction of the magnetic field magnitude, corresponding to a decrease of the vortex density.

\section{Results}
Figure~\ref{DifPatDecay} shows SANS VL diffraction patterns illustrating the main result of this report.
In all instances a pair of Bragg peaks are observed, split by an angle $\omega$ around the crystalline $\textbf{a}$ axis as indicated in Fig.~\ref{DifPatDecay}(a).
A total of six such pairs exist, but to conserve beam time only the peaks at the top of the detector were brought into the Bragg condition.
The split peaks correspond to VL domains rotated about the crystalline $\textbf{c}$ axis in a clockwise or counter clockwise direction, with the intensity difference being due to an unequal domain population~\cite{Avers:2020wx,Avers:2022gh}.
All diffraction patterns in Fig.~\ref{DifPatDecay} were recorded at the same temperature and magnetic field and in the absence of a periodic field oscillation, and show a clear reduction of the intensity as a function of time.
This reflects a gradual VL disordering and a corresponding broadening of the Bragg peaks in reciprocal space. 
Although the poor resolution within the detector plane makes the broadening difficult to resolve~\cite{Muhlbauer:2019jt}, it is clearly seen in the VL rocking curves discussed later.
Note that what is characterized as an ordered VL is most likely a Bragg glass phase with algebraically decaying correlations~\cite{Cubitt:1993aa,Giamarchi:1995tq,Giamarchi:1997th,Klein:2001aa,Divakar:2004by,Laver:2008aa,ToftPetersen:2018ch}, although it is not possible to establish this conclusively from the present SANS data.

\begin{figure}[b]
	\includegraphics[scale = 0.9]{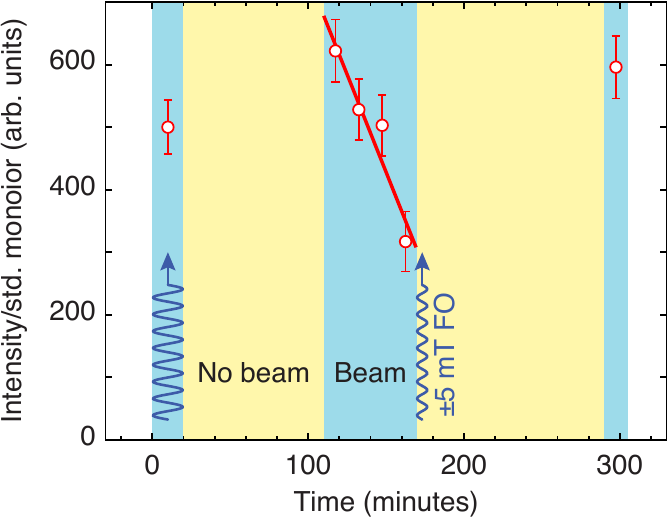}
	\caption{\label{ShutterTest}
	    Exposure test on ZR11 at 0.8~T (ORNL), where dark (light) shading indicate when the neutron beam was on (off).
	    The VL scattered intensity is normalized to the standard monitor count.
	    Field oscillations (FO) of $\pm 5$~mT were applied periodically during the first 20 minute measurement and at 170 minutes after the neutron beam was shut off.
	    }
\end{figure}

To verify that the VL disordering is due to the neutron beam, measurements were performed both during a continuous exposure and following a prolonged period with the beam turned off.
These are summarized in Fig.~\ref{ShutterTest}.
\begin{figure*}
	\includegraphics[width = \linewidth]{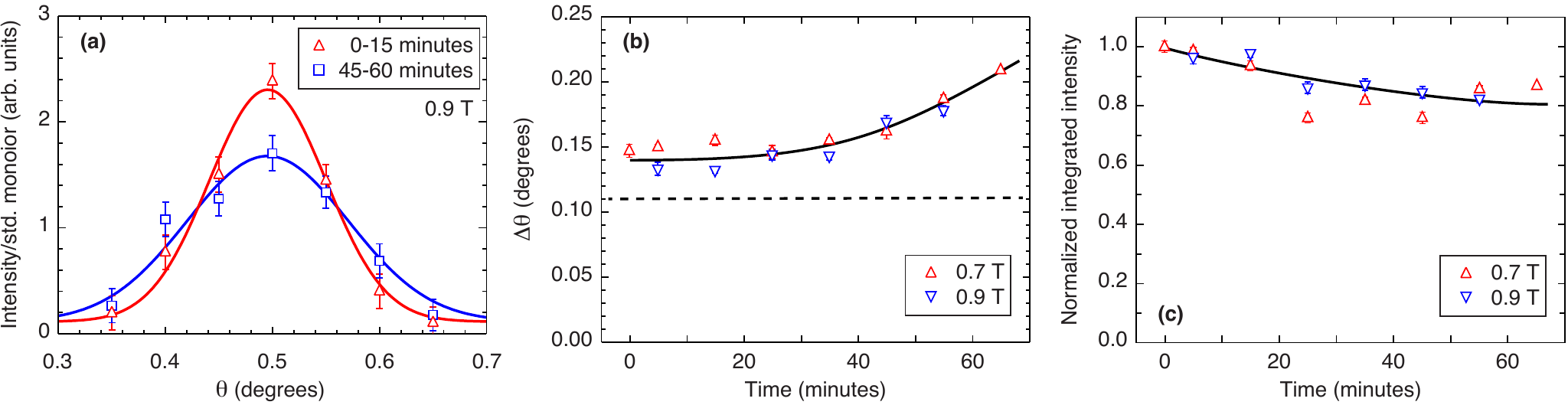}
	\caption{\label{RC}
	    (a) Rocking curves obtained on ZR8 at 0.9~T (ILL), measured during the first 15 minutes after the preparation of a pristine VL and after 45-60 minutes.
	    Curves are Gaussian fits to the data as described in the text.
	    (b) Time dependence of the fitted rocking curve width.
	    The dotted line indicates the experimental resolution.
	    (c) Time dependence of the fitted integrated intensity, normalized to the value for the pristine VL.
	    Solid lines in (b) and (c) are guides to the eye.
	    }
\end{figure*}
Prior to the measurements a pristine VL was prepared, and the first data point (0-20 minutes) was collected while a $\pm 5$~mT field oscillation was performed roughly every 60 seconds to maintain an ordered VL.
The neutron shutter was then closed and the periodic field oscillation was turned off.
After 90 minutes the shutter was re-opened and four subsequent 15 minute measurements of the VL intensity were made.
The shutter was then closed again and a $\pm 5$~mT field oscillation was applied to re-anneal the VL.
After an additional two hours, the shutter was opened for a final 15 minute measurement.

The data in Fig.~\ref{ShutterTest} establish conclusively that the VL disordering occurs when the sample is illuminated by the neutron beam, and is attributed to $^{235}$U fission events in the sample.
Specifically, the intensity decreases to roughly half its initial value between 110 and 170 minutes, while no reduction is observed during the two periods where the beam was off.
Figure~\ref{ShutterTest} also confirms that an ordered VL is achieved by the application of the damped field oscillation, indicated by the recovery of the intensity both in the second and in the final data point.
Finally, the intensity measured after the shutter opening at 110 and 290 minutes is somewhat higher than for the first data point.
This is ascribed to the prolonged absence of fission heating of the {\UPt} crystals resulting in a lower overall sample temperature (see Supplemental Material).
The possibility of a spontaneous re-ordering of the VL was also investigated.
After one hour without beam exposure the intensity of a disordered VL was found to increase only modestly, with a count rate within measurement error of that recorded immediately before the beam was turned off.
Again, the slight increase is likely due due to a lower sample temperature, and any re-ordering thus occurs on time scales much longer than the beam induced disordering if at all.

We now return to the broadening of the VL reflections perpendicular to the detector plane.
Figure~\ref{RC}(a) shows rocking curves of the scattered intensity at 0.9~T as the split VL peak is rotated through the Bragg condition, where $\theta$ is the angle between the magnetic field and the neutron beam direction.
An increase in the width and a reduction of the maximum intensity at $\theta \approx 0.5^{\circ}$ is clearly observed at the later time, consistent with Figs.~\ref{DifPatDecay} and \ref{ShutterTest} which were measured a the peak of the rocking curve.
The rocking curves are fitted to a Gaussian
\begin{equation}
    I(\theta) = \frac{I_{\text{VL}}}{\Delta \theta} \; \exp \left[ -2 \ln (4) \left( \frac{\theta-\theta_0}{\Delta \theta} \right)^2 \right],
    \label{RCfit}
\end{equation}
where $I_{\text{VL}}$ is proportional to the integrated intensity, $\theta_0$ is the peak position and $\Delta \theta$ is the full width half maximum (FWHM).
Fitted values of $I_{\text{VL}}$ and $\Delta \theta$ as a function of time for fields of 0.7~T and 0.9~T are shown in Figs.~\ref{RC}(b) and \ref{RC}(c) respectively.
The integrated intensity is normalized to the value for the pristine VL, obtained either from measurements performed in the presence of a periodic $\pm 5$~mT field oscillation (0.7~T) or by a linear extrapolation of  $I_{\text{VL}}$ to zero time (0.9~T).

The rocking curve width increases with time for both measured fields as seen in Fig.~\ref{RC}(b), confirming the gradual VL disordering hypothesized earlier.
At shorter times $\Delta \theta$ approach the experimental resolution estimated by
\begin{equation}
    \Delta \theta_{\text{res.}} = \sqrt{\delta \theta^2 + \left( \frac{q_{\text{VL}} \lambda_{\text{n}}}{2 \pi} \frac{\Delta \lambda_{\text{n}}}{\lambda_{\text{n}}} \right)^2},
    \label{RCres}
\end{equation}
where $\delta \theta$ is the standard deviation of the beam divergence, $\lambda_{\text{n}} = 0.75$~nm is the neutron wavelength, and $\Delta \lambda_{\text{n}}/\lambda_{\text{n}} = 15\%$ (ORNL) or 10\% (ILL) is the FWHM wavelength spread~\cite{Pedersen:1990aa}.
The VL scattering vector $q_{\text{VL}} = 2\pi (2B/\sqrt{3} \phi_0)^{1/2}$, where $\phi_0 = 2068$~T~nm$^2$ is the flux quantum.
The near constant value of $\Delta \theta$ during the first roughly 30~minutes indicates an intrinsic VL Bragg peaks width which is much smaller than $\Delta \theta_{\text{res.}}$, with a clear broadening observed only once the two becomes comparable.
The reciprocal rocking curve width (in radians) can thus be taken as a lower bound on the longitudinal VL correlation length, $\zeta_L \geq 2(q_{\text{VL}} \, \Delta \theta)^{-1} \approx 6-7$~$\mu$m.
The absence of a time lag in reduction of the integrated intensity in Fig.~\ref{RC}(c) provides further evidence that the VL disordering begins as soon as the sample is subjected to the neutron beam.
Here, $I_{\text{VL}}$ decreases by $\sim 20\%$ within the first hour, possibly leveling off at the longer times.
The decrease of $I_{\text{VL}}$ indicates a disruption of the VL, and possibly an evolution towards a vortex glass phase~\cite{ToftPetersen:2018ch}.
Notably, the 0.9~T measurements were not preceded by a beam-off period and the reduction of $I_{\text{VL}}$ is thus not due to a gradual heating of the sample.

\begin{figure}
	\includegraphics[scale = 0.9]{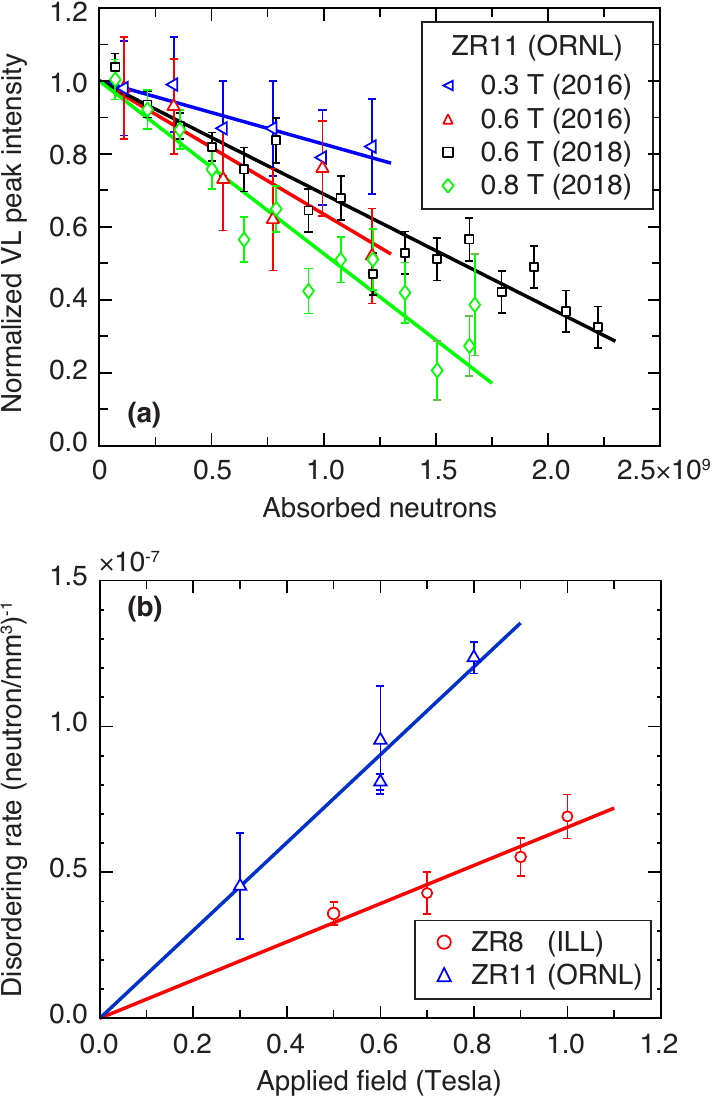}
	\caption{\label{DisorderingRate}
	    (a) VL scattering rate vs absorbed neutrons per area transmitted through the sample for different fields.
	    Data were obtained on ZR11 at ORNL during separate experiments (2016 and 2018).
	    Intensities are normalized such that linear fits (solid lines) extrapolate to unity for the pristine VL.
	    (b) Disordering rate per unit volume vs magnetic field for both {\UPt} samples.
	    Lines are linear fits to the ZR8 and ZR11 data, constrained to pass through the origin.
	    }
\end{figure}

To quantify the VL disordering, measurements of the rocking curve peak intensity as a function of neutron exposure were performed for a range of magnetic fields.
Figure~\ref{DisorderingRate}(a) shows examples of decay curves for three different fields versus the number of absorbed neutrons, demonstrating that the VL disorders more quickly at higher magnetic fields.
The disordering rates, defined by the slope of linear fits to the data, are summarized in Fig.~\ref{DisorderingRate}(b) for all measurement sequences and both {\UPt} crystals.
Here, the rates are normalized to absorbed neutrons per unit volume to allow for a direct comparison of the ZR8 and ZR11 samples (see Supplementary Material).

While the decreasing intensity in Fig.~\ref{DisorderingRate}(a) could in principle be due to a gradual heating of the sample, several factors allows us to rule this out.
These include the constant VL splitting in Fig.~\ref{DifPatDecay}, which is in contrast to the rapid decrease observed with increasing temperature~\cite{Avers:2022gh}.
Furthermore, successive measurements at 0.6~T, separated only by the application of a damped field oscillation, shows the same decay of the VL intensity (see Supplemental Material).

\section{Discussion}
A strong temperature history-dependence of the {\UPt} VL is well documented.
For example, the ability to thermally quench the VL in {\UPt} was used to facilitate SANS studies of the superconducting A phase which exists just below $T_{\rm c}$~\cite{Huxley:2000aa}.
Here, VL Bragg peaks were observed oriented along or split around the crystalline $\bm{a}$ axis, depending on the quench temperature.
In comparison, a slow cooling from temperatures above $T_{\rm c}$ leads to a different VL configuration, with Bragg peaks along the $\bm{a}^*$ axis.
We do not observe any VL scattering around $\bm{a}^*$, neither in the rocked on measurements or the rocking curves in Fig.~\ref{RC} which would satisfy the Bragg condition for peaks at this location.
The vanishing scattering in our SANS measurements is therefore not due to intensity being transferred from one VL peak to another.

We propose that the observed decreasing intensity with increasing neutron exposure arise from a rapid thermal cycling, due to fission events which locally heat the sample above the critical temperature.
Following the thermal transient  vortices re-form in a disordered configuration due to the quench through the vortex glass state below the superconducting transition~\cite{Blatter:1994gz,ToftPetersen:2018ch}.
Support for such a scenario comes from SANS studies of NbSe$_2$, where the degree of VL ordering was found to depend sensitively on the thermal history in the vicinity of the so-called peak effect in the critical current~\cite{Daniilidis:2007wa,MarzialiBermudez:2015uj,MarzialiBermudez:2017vl}.
The peak effect is associated with the order-disorder transition between the vortex glass and Bragg glass states~\cite{Gammel:1998wj,Joumard:1999aa}, and is also observed in {\UPt}~\cite{Yaron:1997wx}.

The volume affected by a single fission event may be estimated from the low-temperature specific heat ($C/T \approx 1.6 T$~J/K$^3$\,mol) and the mass density ($\rho = 19.4$~g/cm$^3$) of {\UPt}~\cite{Joynt:2002wt}, yielding a sphere of diameter $\sim 34$~$\mu$m heated above $T_{\rm c}$.
This should be considered an order of magnitude estimate since it assumes the applicability of the equilibrium specific heat during the fission process, and ignores both the field dependence of the specific heat~\cite{Ramirez:1995fh} as well as the thermal conductivity~\cite{Suderow:1997ex}.
The size of the affected volume exceeds the 4--6 $\mu$m range of the Ba and Kr fission products in \UPt~\cite{SRIM}.
It is also much greater than the vortex spacing given by $a_0 = (2 \phi_0/\sqrt{3} B)^{1/2}$, with the diameter obtained above corresponding to $\sim 400a_0 - 700 a_0$ as the field is increased from 0.3 to 1~T.
A single fission event thus affects large sections of the VL, several times greater than the longitudinal correlation length estimated from the rocking curves.
 
A detailed understanding of the fission induced VL disordering will require a careful analysis, and is outside the scope of this report.
Nevertheless, two important points may be deduced from Fig.~\ref{DisorderingRate}(b).
First, for each {\UPt} crystal the disordering rate is directly proportional to the applied magnetic field, indicating that it is governed by VL properties.
Here, the elastic moduli, which increase with rising vortex density~\cite{Brandt:1995aa}, will be important to consider as a greater stiffness will allow VL ``shock waves'' to propagate further from the volume directly affected by the fission events.
Second, the ratio of the disordering rate to applied field differs for ZR8 and ZR11.
The latter crystal is of higher quality (see Supplemental Material) and therefore expected to exhibit less vortex pinning.
This will make the ZR11 VL less resilient against disordering, as disruptions can more easily propagate away from the volume directly affected by the fission event.
As stacking faults are the principal crystalline defects in {\UPt}~\cite{Hong:1999ud} and directly visible in the small-angle background scattering~\cite{Huxley:2000aa}, the relation between the disordering rate and sample quality should be quantified in future SANS experiments.

A remaining question is how, and to which degree, the {\UPt} crystals are affected by the fission processes?
The incorporation of fissionable elements and subsequent irradiation by neutron has previously been used to create defects and enhance the vortex pinning in conventional superconductors as well a high-$T_c$ oxides~\cite{Bean:1966dv,Fleischer:1989gd,Shan:2003ds}.
Here, an increase in the critical current was found for fission event densities of the order $10^{14} - 10^{15}$~cm$^{-3}$.
In comparison, the {\UPt} ZR8 crystal has experienced $\sim 10^{11}$ fission events during the roughly one month of accumulated irradiation.
It is therefore not surprising that the intrinsic vortex pinning has not increased sufficiently to affect the SANS measurements, as it is still possible to achieve an ordered VL by a field oscillation.

Before concluding we also consider how the fission events affect the superconducting order parameter in {\UPt} and our recent SANS experiments which provide direct evidence for broken time reversal symmetry in the B phase~\cite{Avers:2020wx}.
Here, differences in the VL orientation relative to the crystalline axes were observed between states with the Cooper pair orbital angular momentum either parallel or anti-parallel to the applied magnetic field.
The latter corresponds to a metastable order parameter configuration, achieved by reducing the magnetic field at low temperature to enter the B phase and establish the chiral direction, eventually passing through zero and thus reverse the field direction.
Fission events could potentially impact the chiral direction in the regions temporarily heated above $T_{\rm c}$, and, if this direction is reversed upon re-cooling, lead to the formation of order parameter domain boundaries.

Based on the rate and affected volume, the fraction of the sample that remains unaffected by fission will decrease exponentially with a time constant of approximately 6~minutes.
This is much shorter than the SANS count times which, at the higher fields, reach several hours.
Since a difference in the VL was indeed observed between the two order parameter configurations~\cite{Avers:2020wx}, a fission induced reversal of the chiral direction, and the accompanying order parameter domain formation, does not appear to take place.
This is consistent with both tunneling~\cite{Strand:2009eq} and Kerr rotation~\cite{Schemm:2014fv} measurements, showing an absence of order parameter domain formation in {\UPt} even in the absence of a training field, and suggesting that a single order parameter domain spans the entire crystal~\cite{Schemm:2014fv}.
Using the roughly 1~mm length scale associated with these experiments, as a lower limit on the domain size yields a volume much greater than that affected by a fission event.
This implies that domain formation in {\UPt} is energetically unfavorable, and supports our conclusion that the order parameter in the SANS measurements is reformed with the same chiral direction as the surrounding sample.

\section{Conclusion}
In summary, we have shown that the VL in {\UPt} undergoes a gradual fission induced disordering when exposed to a beam of cold neutrons.
Fission events heat regions of the sample above $T_{\rm c}$ which upon re-cooling hosts a quenched vortex glass state.
The disordering rate is proportional to the vortex density, suggesting that it is governed by the VL stiffness. 
Our observations provide new possibilities for studies of vortex matter whereby local and reversible disorder can be combined with e.g. thermal disordering. 
We also speculate that the magnitude of field oscillation required for maintaining an ordered VL will be less than that needed to re-order a disordered VL, due to the different structural properties.
Finally, the use of periodic field oscillations to mitigate the VL disordering can be applied to other U-containing superconductors where SANS studies have to date been unsuccessful.

\begin{acknowledgments}
We are grateful to J.~A.~Sauls for numerous discussions over the years and to U.~K\"{o}ster for providing estimates of the sample heating due to fission.
This work was supported by the U.S. Department of Energy, Office of Basic Energy Sciences, under Awards No.~DE-SC0005051 (MRE: University of Notre Dame; neutron scattering) and DE-FG02-05ER46248 (WPH: Northwestern University; crystal growth and neutron scattering).
A portion of this research used resources at the High Flux Isotope Reactor, a DOE Office of Science User Facility operated by the Oak Ridge National Laboratory.
Part of this work is based on experiments performed at the Institut Laue-Langevin, Grenoble, France.
\end{acknowledgments}

\bibliography{UPt3Fission}

\begin{thebibliography}{58}%
\makeatletter
\providecommand \@ifxundefined [1]{%
 \@ifx{#1\undefined}
}%
\providecommand \@ifnum [1]{%
 \ifnum #1\expandafter \@firstoftwo
 \else \expandafter \@secondoftwo
 \fi
}%
\providecommand \@ifx [1]{%
 \ifx #1\expandafter \@firstoftwo
 \else \expandafter \@secondoftwo
 \fi
}%
\providecommand \natexlab [1]{#1}%
\providecommand \enquote  [1]{``#1''}%
\providecommand \bibnamefont  [1]{#1}%
\providecommand \bibfnamefont [1]{#1}%
\providecommand \citenamefont [1]{#1}%
\providecommand \href@noop [0]{\@secondoftwo}%
\providecommand \href [0]{\begingroup \@sanitize@url \@href}%
\providecommand \@href[1]{\@@startlink{#1}\@@href}%
\providecommand \@@href[1]{\endgroup#1\@@endlink}%
\providecommand \@sanitize@url [0]{\catcode `\\12\catcode `\$12\catcode
  `\&12\catcode `\#12\catcode `\^12\catcode `\_12\catcode `\%12\relax}%
\providecommand \@@startlink[1]{}%
\providecommand \@@endlink[0]{}%
\providecommand \url  [0]{\begingroup\@sanitize@url \@url }%
\providecommand \@url [1]{\endgroup\@href {#1}{\urlprefix }}%
\providecommand \urlprefix  [0]{URL }%
\providecommand \Eprint [0]{\href }%
\providecommand \doibase [0]{http://dx.doi.org/}%
\providecommand \selectlanguage [0]{\@gobble}%
\providecommand \bibinfo  [0]{\@secondoftwo}%
\providecommand \bibfield  [0]{\@secondoftwo}%
\providecommand \translation [1]{[#1]}%
\providecommand \BibitemOpen [0]{}%
\providecommand \bibitemStop [0]{}%
\providecommand \bibitemNoStop [0]{.\EOS\space}%
\providecommand \EOS [0]{\spacefactor3000\relax}%
\providecommand \BibitemShut  [1]{\csname bibitem#1\endcsname}%
\let\auto@bib@innerbib\@empty
\bibitem [{\citenamefont {Huebener}(2001)}]{Huebener:2001we}%
  \BibitemOpen
  \bibfield  {author} {\bibinfo {author} {\bibfnamefont {R.~P.}\ \bibnamefont
  {Huebener}},\ }\href@noop {} {\emph {\bibinfo {title} {{Magnetic Flux
  Structures in Superconductors}}}}\ (\bibinfo  {publisher} {Springer},\
  \bibinfo {year} {2001})\BibitemShut {NoStop}%
\bibitem [{\citenamefont {Abrikosov}(1957)}]{Abrikosov:1957vu}%
  \BibitemOpen
  \bibfield  {author} {\bibinfo {author} {\bibfnamefont {A.~A.}\ \bibnamefont
  {Abrikosov}},\ }\href@noop {} {\bibfield  {journal} {\bibinfo  {journal}
  {Sov. Phys. JETP}\ }\textbf {\bibinfo {volume} {5}},\ \bibinfo {pages} {1174}
  (\bibinfo {year} {1957})}\BibitemShut {NoStop}%
\bibitem [{\citenamefont {Kleiner}\ \emph {et~al.}(1964)\citenamefont
  {Kleiner}, \citenamefont {Autler},\ and\ \citenamefont
  {Roth}}]{Kleiner:1964ih}%
  \BibitemOpen
  \bibfield  {author} {\bibinfo {author} {\bibfnamefont {W.~H.}\ \bibnamefont
  {Kleiner}}, \bibinfo {author} {\bibfnamefont {S.~H.}\ \bibnamefont {Autler}},
  \ and\ \bibinfo {author} {\bibfnamefont {L.~M.}\ \bibnamefont {Roth}},\
  }\href@noop {} {\bibfield  {journal} {\bibinfo  {journal} {Phys. Rev.}\
  }\textbf {\bibinfo {volume} {133}},\ \bibinfo {pages} {1226} (\bibinfo {year}
  {1964})}\BibitemShut {NoStop}%
\bibitem [{\citenamefont {Matricon}(1964)}]{Matricon:1964bt}%
  \BibitemOpen
  \bibfield  {author} {\bibinfo {author} {\bibfnamefont {J.}~\bibnamefont
  {Matricon}},\ }\href@noop {} {\bibfield  {journal} {\bibinfo  {journal}
  {Phys. Lett.}\ }\textbf {\bibinfo {volume} {9}},\ \bibinfo {pages} {289}
  (\bibinfo {year} {1964})}\BibitemShut {NoStop}%
\bibitem [{\citenamefont {Blatter}\ \emph {et~al.}(1994)\citenamefont
  {Blatter}, \citenamefont {Feigel'man}, \citenamefont {Geshkenbein},
  \citenamefont {Larkin},\ and\ \citenamefont {Vinokur}}]{Blatter:1994gz}%
  \BibitemOpen
  \bibfield  {author} {\bibinfo {author} {\bibfnamefont {G.}~\bibnamefont
  {Blatter}}, \bibinfo {author} {\bibfnamefont {M.~V.}\ \bibnamefont
  {Feigel'man}}, \bibinfo {author} {\bibfnamefont {V.~B.}\ \bibnamefont
  {Geshkenbein}}, \bibinfo {author} {\bibfnamefont {A.~I.}\ \bibnamefont
  {Larkin}}, \ and\ \bibinfo {author} {\bibfnamefont {V.~M.}\ \bibnamefont
  {Vinokur}},\ }\href@noop {} {\bibfield  {journal} {\bibinfo  {journal} {Rev.
  Mod. Phys.}\ }\textbf {\bibinfo {volume} {66}},\ \bibinfo {pages} {1125}
  (\bibinfo {year} {1994})}\BibitemShut {NoStop}%
\bibitem [{\citenamefont {Brandt}(1995)}]{Brandt:1995aa}%
  \BibitemOpen
  \bibfield  {author} {\bibinfo {author} {\bibfnamefont {E.~H.}\ \bibnamefont
  {Brandt}},\ }\href@noop {} {\bibfield  {journal} {\bibinfo  {journal} {Rep.
  Prog. Phys.}\ }\textbf {\bibinfo {volume} {58}},\ \bibinfo {pages} {1465}
  (\bibinfo {year} {1995})}\BibitemShut {NoStop}%
\bibitem [{\citenamefont {Giamarchi}\ and\ \citenamefont
  {Le~Doussal}(1995)}]{Giamarchi:1995tq}%
  \BibitemOpen
  \bibfield  {author} {\bibinfo {author} {\bibfnamefont {T.}~\bibnamefont
  {Giamarchi}}\ and\ \bibinfo {author} {\bibfnamefont {P.}~\bibnamefont
  {Le~Doussal}},\ }\href@noop {} {\bibfield  {journal} {\bibinfo  {journal}
  {Phys. Rev. B}\ }\textbf {\bibinfo {volume} {52}},\ \bibinfo {pages} {1242}
  (\bibinfo {year} {1995})}\BibitemShut {NoStop}%
\bibitem [{\citenamefont {Giamarchi}\ and\ \citenamefont
  {Le~Doussal}(1997)}]{Giamarchi:1997th}%
  \BibitemOpen
  \bibfield  {author} {\bibinfo {author} {\bibfnamefont {T.}~\bibnamefont
  {Giamarchi}}\ and\ \bibinfo {author} {\bibfnamefont {P.}~\bibnamefont
  {Le~Doussal}},\ }\href@noop {} {\bibfield  {journal} {\bibinfo  {journal}
  {Phys. Rev. B}\ }\textbf {\bibinfo {volume} {55}},\ \bibinfo {pages} {6577}
  (\bibinfo {year} {1997})}\BibitemShut {NoStop}%
\bibitem [{\citenamefont {Nattermann}\ and\ \citenamefont
  {Scheidl}(2000)}]{Nattermann:2000vk}%
  \BibitemOpen
  \bibfield  {author} {\bibinfo {author} {\bibfnamefont {T.}~\bibnamefont
  {Nattermann}}\ and\ \bibinfo {author} {\bibfnamefont {S.}~\bibnamefont
  {Scheidl}},\ }\href@noop {} {\bibfield  {journal} {\bibinfo  {journal} {Adv.
  Phys.}\ }\textbf {\bibinfo {volume} {49}},\ \bibinfo {pages} {607} (\bibinfo
  {year} {2000})}\BibitemShut {NoStop}%
\bibitem [{\citenamefont {Le~Doussal}(2010)}]{LeDoussal:2010wu}%
  \BibitemOpen
  \bibfield  {author} {\bibinfo {author} {\bibfnamefont {P.}~\bibnamefont
  {Le~Doussal}},\ }\href@noop {} {\bibfield  {journal} {\bibinfo  {journal}
  {Int. J. Mod. Phys. B}\ }\textbf {\bibinfo {volume} {24}},\ \bibinfo {pages}
  {3855} (\bibinfo {year} {2010})}\BibitemShut {NoStop}%
\bibitem [{\citenamefont {Menghini}\ \emph {et~al.}(2003)\citenamefont
  {Menghini}, \citenamefont {Fasano}, \citenamefont {de~la Cruz}, \citenamefont
  {Banerjee}, \citenamefont {Myasoedov}, \citenamefont {Zeldov}, \citenamefont
  {van~der Beek}, \citenamefont {Konczykowski},\ and\ \citenamefont
  {Tamegai}}]{Menghini:2003fe}%
  \BibitemOpen
  \bibfield  {author} {\bibinfo {author} {\bibfnamefont {M.}~\bibnamefont
  {Menghini}}, \bibinfo {author} {\bibfnamefont {Y.}~\bibnamefont {Fasano}},
  \bibinfo {author} {\bibfnamefont {F.}~\bibnamefont {de~la Cruz}}, \bibinfo
  {author} {\bibfnamefont {S.~S.}\ \bibnamefont {Banerjee}}, \bibinfo {author}
  {\bibfnamefont {Y.}~\bibnamefont {Myasoedov}}, \bibinfo {author}
  {\bibfnamefont {E.}~\bibnamefont {Zeldov}}, \bibinfo {author} {\bibfnamefont
  {C.~J.}\ \bibnamefont {van~der Beek}}, \bibinfo {author} {\bibfnamefont
  {M.}~\bibnamefont {Konczykowski}}, \ and\ \bibinfo {author} {\bibfnamefont
  {T.}~\bibnamefont {Tamegai}},\ }\href@noop {} {\bibfield  {journal} {\bibinfo
   {journal} {Phys. Rev. Lett.}\ }\textbf {\bibinfo {volume} {90}},\ \bibinfo
  {pages} {147001} (\bibinfo {year} {2003})}\BibitemShut {NoStop}%
\bibitem [{\citenamefont {Checchin}\ and\ \citenamefont
  {Grassellino}(2020)}]{Checchin:2020ko}%
  \BibitemOpen
  \bibfield  {author} {\bibinfo {author} {\bibfnamefont {M.}~\bibnamefont
  {Checchin}}\ and\ \bibinfo {author} {\bibfnamefont {A.}~\bibnamefont
  {Grassellino}},\ }\href@noop {} {\bibfield  {journal} {\bibinfo  {journal}
  {Phys. Rev. Appl.}\ }\textbf {\bibinfo {volume} {14}},\ \bibinfo {pages}
  {044018} (\bibinfo {year} {2020})}\BibitemShut {NoStop}%
\bibitem [{\citenamefont {Romanenko}\ \emph {et~al.}(2020)\citenamefont
  {Romanenko}, \citenamefont {Pilipenko}, \citenamefont {Zorzetti},
  \citenamefont {Frolov}, \citenamefont {Awida}, \citenamefont {Belomestnykh},
  \citenamefont {Posen},\ and\ \citenamefont {Grassellino}}]{Romanenko:2020jj}%
  \BibitemOpen
  \bibfield  {author} {\bibinfo {author} {\bibfnamefont {A.}~\bibnamefont
  {Romanenko}}, \bibinfo {author} {\bibfnamefont {R.}~\bibnamefont
  {Pilipenko}}, \bibinfo {author} {\bibfnamefont {S.}~\bibnamefont {Zorzetti}},
  \bibinfo {author} {\bibfnamefont {D.}~\bibnamefont {Frolov}}, \bibinfo
  {author} {\bibfnamefont {M.}~\bibnamefont {Awida}}, \bibinfo {author}
  {\bibfnamefont {S.}~\bibnamefont {Belomestnykh}}, \bibinfo {author}
  {\bibfnamefont {S.}~\bibnamefont {Posen}}, \ and\ \bibinfo {author}
  {\bibfnamefont {A.}~\bibnamefont {Grassellino}},\ }\href@noop {} {\bibfield
  {journal} {\bibinfo  {journal} {Phys. Rev. Appl.}\ }\textbf {\bibinfo
  {volume} {13}},\ \bibinfo {pages} {034032} (\bibinfo {year}
  {2020})}\BibitemShut {NoStop}%
\bibitem [{\citenamefont {Xiao}\ \emph {et~al.}(2004)\citenamefont {Xiao},
  \citenamefont {Dogru}, \citenamefont {Andrei}, \citenamefont {Shuk},\ and\
  \citenamefont {Greenblatt}}]{Xiao:2004gk}%
  \BibitemOpen
  \bibfield  {author} {\bibinfo {author} {\bibfnamefont {Z.~L.}\ \bibnamefont
  {Xiao}}, \bibinfo {author} {\bibfnamefont {O.}~\bibnamefont {Dogru}},
  \bibinfo {author} {\bibfnamefont {E.~Y.}\ \bibnamefont {Andrei}}, \bibinfo
  {author} {\bibfnamefont {P.}~\bibnamefont {Shuk}}, \ and\ \bibinfo {author}
  {\bibfnamefont {M.}~\bibnamefont {Greenblatt}},\ }\href@noop {} {\bibfield
  {journal} {\bibinfo  {journal} {Phys. Rev. Lett.}\ }\textbf {\bibinfo
  {volume} {92}},\ \bibinfo {pages} {227004} (\bibinfo {year}
  {2004})}\BibitemShut {NoStop}%
\bibitem [{\citenamefont {Marziali~Bermudez}\ \emph {et~al.}(2015)\citenamefont
  {Marziali~Bermudez}, \citenamefont {Eskildsen}, \citenamefont {Bartkowiak},
  \citenamefont {Nagy}, \citenamefont {Bekeris},\ and\ \citenamefont
  {Pasquini}}]{MarzialiBermudez:2015uj}%
  \BibitemOpen
  \bibfield  {author} {\bibinfo {author} {\bibfnamefont {M.}~\bibnamefont
  {Marziali~Bermudez}}, \bibinfo {author} {\bibfnamefont {M.~R.}\ \bibnamefont
  {Eskildsen}}, \bibinfo {author} {\bibfnamefont {M.}~\bibnamefont
  {Bartkowiak}}, \bibinfo {author} {\bibfnamefont {G.}~\bibnamefont {Nagy}},
  \bibinfo {author} {\bibfnamefont {V.}~\bibnamefont {Bekeris}}, \ and\
  \bibinfo {author} {\bibfnamefont {G.}~\bibnamefont {Pasquini}},\ }\href@noop
  {} {\bibfield  {journal} {\bibinfo  {journal} {Phys. Rev. Lett.}\ }\textbf
  {\bibinfo {volume} {115}},\ \bibinfo {pages} {067001} (\bibinfo {year}
  {2015})}\BibitemShut {NoStop}%
\bibitem [{\citenamefont {Marziali~Bermudez}\ \emph {et~al.}(2017)\citenamefont
  {Marziali~Bermudez}, \citenamefont {Louden}, \citenamefont {Eskildsen},
  \citenamefont {Dewhurst}, \citenamefont {Bekeris},\ and\ \citenamefont
  {Pasquini}}]{MarzialiBermudez:2017vl}%
  \BibitemOpen
  \bibfield  {author} {\bibinfo {author} {\bibfnamefont {M.}~\bibnamefont
  {Marziali~Bermudez}}, \bibinfo {author} {\bibfnamefont {E.~R.}\ \bibnamefont
  {Louden}}, \bibinfo {author} {\bibfnamefont {M.~R.}\ \bibnamefont
  {Eskildsen}}, \bibinfo {author} {\bibfnamefont {C.~D.}\ \bibnamefont
  {Dewhurst}}, \bibinfo {author} {\bibfnamefont {V.}~\bibnamefont {Bekeris}}, \
  and\ \bibinfo {author} {\bibfnamefont {G.}~\bibnamefont {Pasquini}},\
  }\href@noop {} {\bibfield  {journal} {\bibinfo  {journal} {Phys. Rev. B}\
  }\textbf {\bibinfo {volume} {95}},\ \bibinfo {pages} {104505} (\bibinfo
  {year} {2017})}\BibitemShut {NoStop}%
\bibitem [{\citenamefont {Huxley}\ \emph {et~al.}(2000)\citenamefont {Huxley},
  \citenamefont {Rodi{\`e}re}, \citenamefont {Paul}, \citenamefont {van Dijk},
  \citenamefont {Cubitt},\ and\ \citenamefont {Flouquet}}]{Huxley:2000aa}%
  \BibitemOpen
  \bibfield  {author} {\bibinfo {author} {\bibfnamefont {A.}~\bibnamefont
  {Huxley}}, \bibinfo {author} {\bibfnamefont {P.}~\bibnamefont {Rodi{\`e}re}},
  \bibinfo {author} {\bibfnamefont {D.~M.}\ \bibnamefont {Paul}}, \bibinfo
  {author} {\bibfnamefont {N.}~\bibnamefont {van Dijk}}, \bibinfo {author}
  {\bibfnamefont {R.}~\bibnamefont {Cubitt}}, \ and\ \bibinfo {author}
  {\bibfnamefont {J.}~\bibnamefont {Flouquet}},\ }\href@noop {} {\bibfield
  {journal} {\bibinfo  {journal} {Nature}\ }\textbf {\bibinfo {volume} {406}},\
  \bibinfo {pages} {160} (\bibinfo {year} {2000})}\BibitemShut {NoStop}%
\bibitem [{\citenamefont {Das}\ \emph {et~al.}(2012)\citenamefont {Das},
  \citenamefont {Rastovski}, \citenamefont {O'Brien}, \citenamefont
  {Schlesinger}, \citenamefont {Dewhurst}, \citenamefont {DeBeer-Schmitt},
  \citenamefont {Zhigadlo}, \citenamefont {Karpinski},\ and\ \citenamefont
  {Eskildsen}}]{Das:2012cf}%
  \BibitemOpen
  \bibfield  {author} {\bibinfo {author} {\bibfnamefont {P.}~\bibnamefont
  {Das}}, \bibinfo {author} {\bibfnamefont {C.}~\bibnamefont {Rastovski}},
  \bibinfo {author} {\bibfnamefont {T.~R.}\ \bibnamefont {O'Brien}}, \bibinfo
  {author} {\bibfnamefont {K.~J.}\ \bibnamefont {Schlesinger}}, \bibinfo
  {author} {\bibfnamefont {C.~D.}\ \bibnamefont {Dewhurst}}, \bibinfo {author}
  {\bibfnamefont {L.}~\bibnamefont {DeBeer-Schmitt}}, \bibinfo {author}
  {\bibfnamefont {N.~D.}\ \bibnamefont {Zhigadlo}}, \bibinfo {author}
  {\bibfnamefont {J.}~\bibnamefont {Karpinski}}, \ and\ \bibinfo {author}
  {\bibfnamefont {M.~R.}\ \bibnamefont {Eskildsen}},\ }\href@noop {} {\bibfield
   {journal} {\bibinfo  {journal} {Phys. Rev. Lett.}\ }\textbf {\bibinfo
  {volume} {108}},\ \bibinfo {pages} {167001} (\bibinfo {year}
  {2012})}\BibitemShut {NoStop}%
\bibitem [{\citenamefont {Okuma}\ \emph {et~al.}(2012)\citenamefont {Okuma},
  \citenamefont {Shimamoto},\ and\ \citenamefont {Kokubo}}]{Okuma:2012dz}%
  \BibitemOpen
  \bibfield  {author} {\bibinfo {author} {\bibfnamefont {S.}~\bibnamefont
  {Okuma}}, \bibinfo {author} {\bibfnamefont {D.}~\bibnamefont {Shimamoto}}, \
  and\ \bibinfo {author} {\bibfnamefont {N.}~\bibnamefont {Kokubo}},\
  }\href@noop {} {\bibfield  {journal} {\bibinfo  {journal} {Phys. Rev. B}\
  }\textbf {\bibinfo {volume} {85}},\ \bibinfo {pages} {064508} (\bibinfo
  {year} {2012})}\BibitemShut {NoStop}%
\bibitem [{\citenamefont {Yaron}\ \emph {et~al.}(1994)\citenamefont {Yaron},
  \citenamefont {Gammel}, \citenamefont {Huse}, \citenamefont {Kleiman},
  \citenamefont {Oglesby}, \citenamefont {Bucher}, \citenamefont {Batlogg},
  \citenamefont {Bishop}, \citenamefont {Mortensen}, \citenamefont {Clausen},
  \citenamefont {Bolle},\ and\ \citenamefont {de~La~Cruz}}]{Yaron:1994tb}%
  \BibitemOpen
  \bibfield  {author} {\bibinfo {author} {\bibfnamefont {U.}~\bibnamefont
  {Yaron}}, \bibinfo {author} {\bibfnamefont {P.~L.}\ \bibnamefont {Gammel}},
  \bibinfo {author} {\bibfnamefont {D.~A.}\ \bibnamefont {Huse}}, \bibinfo
  {author} {\bibfnamefont {R.~N.}\ \bibnamefont {Kleiman}}, \bibinfo {author}
  {\bibfnamefont {C.~S.}\ \bibnamefont {Oglesby}}, \bibinfo {author}
  {\bibfnamefont {E.}~\bibnamefont {Bucher}}, \bibinfo {author} {\bibfnamefont
  {B.}~\bibnamefont {Batlogg}}, \bibinfo {author} {\bibfnamefont {D.~J.}\
  \bibnamefont {Bishop}}, \bibinfo {author} {\bibfnamefont {K.}~\bibnamefont
  {Mortensen}}, \bibinfo {author} {\bibfnamefont {K.}~\bibnamefont {Clausen}},
  \bibinfo {author} {\bibfnamefont {C.~A.}\ \bibnamefont {Bolle}}, \ and\
  \bibinfo {author} {\bibfnamefont {F.}~\bibnamefont {de~La~Cruz}},\
  }\href@noop {} {\bibfield  {journal} {\bibinfo  {journal} {Phys. Rev. Lett.}\
  }\textbf {\bibinfo {volume} {73}},\ \bibinfo {pages} {2748} (\bibinfo {year}
  {1994})}\BibitemShut {NoStop}%
\bibitem [{\citenamefont {Yaron}\ \emph {et~al.}(1995)\citenamefont {Yaron},
  \citenamefont {Gammel}, \citenamefont {Huse}, \citenamefont {Kleiman},
  \citenamefont {Oglesby}, \citenamefont {Bucher}, \citenamefont {Batlogg},
  \citenamefont {Bishop}, \citenamefont {Mortensen},\ and\ \citenamefont
  {Clausen}}]{Yaron:1995wa}%
  \BibitemOpen
  \bibfield  {author} {\bibinfo {author} {\bibfnamefont {U.}~\bibnamefont
  {Yaron}}, \bibinfo {author} {\bibfnamefont {P.~L.}\ \bibnamefont {Gammel}},
  \bibinfo {author} {\bibfnamefont {D.~A.}\ \bibnamefont {Huse}}, \bibinfo
  {author} {\bibfnamefont {R.~N.}\ \bibnamefont {Kleiman}}, \bibinfo {author}
  {\bibfnamefont {C.~S.}\ \bibnamefont {Oglesby}}, \bibinfo {author}
  {\bibfnamefont {E.}~\bibnamefont {Bucher}}, \bibinfo {author} {\bibfnamefont
  {B.}~\bibnamefont {Batlogg}}, \bibinfo {author} {\bibfnamefont {D.~J.}\
  \bibnamefont {Bishop}}, \bibinfo {author} {\bibfnamefont {K.}~\bibnamefont
  {Mortensen}}, \ and\ \bibinfo {author} {\bibfnamefont {K.~N.}\ \bibnamefont
  {Clausen}},\ }\href@noop {} {\bibfield  {journal} {\bibinfo  {journal}
  {Nature}\ }\textbf {\bibinfo {volume} {376}},\ \bibinfo {pages} {753}
  (\bibinfo {year} {1995})}\BibitemShut {NoStop}%
\bibitem [{\citenamefont {Duarte}\ \emph {et~al.}(1996)\citenamefont {Duarte},
  \citenamefont {Fernandez~Righi}, \citenamefont {Bolle}, \citenamefont {de~la
  Cruz}, \citenamefont {Gammel}, \citenamefont {Oglesby}, \citenamefont
  {Bucher}, \citenamefont {Batlogg},\ and\ \citenamefont
  {Bishop}}]{Duarte:1996gl}%
  \BibitemOpen
  \bibfield  {author} {\bibinfo {author} {\bibfnamefont {A.}~\bibnamefont
  {Duarte}}, \bibinfo {author} {\bibfnamefont {E.}~\bibnamefont
  {Fernandez~Righi}}, \bibinfo {author} {\bibfnamefont {C.~A.}\ \bibnamefont
  {Bolle}}, \bibinfo {author} {\bibfnamefont {F.}~\bibnamefont {de~la Cruz}},
  \bibinfo {author} {\bibfnamefont {P.~L.}\ \bibnamefont {Gammel}}, \bibinfo
  {author} {\bibfnamefont {C.~S.}\ \bibnamefont {Oglesby}}, \bibinfo {author}
  {\bibfnamefont {E.}~\bibnamefont {Bucher}}, \bibinfo {author} {\bibfnamefont
  {B.}~\bibnamefont {Batlogg}}, \ and\ \bibinfo {author} {\bibfnamefont
  {D.~J.}\ \bibnamefont {Bishop}},\ }\href@noop {} {\bibfield  {journal}
  {\bibinfo  {journal} {Phys. Rev. B}\ }\textbf {\bibinfo {volume} {53}},\
  \bibinfo {pages} {11336} (\bibinfo {year} {1996})}\BibitemShut {NoStop}%
\bibitem [{\citenamefont {Pautrat}\ \emph {et~al.}(2005)\citenamefont
  {Pautrat}, \citenamefont {Scola}, \citenamefont {Simon}, \citenamefont
  {Br{\^u}let}, \citenamefont {Goupil}, \citenamefont {Higgins},\ and\
  \citenamefont {Bhattacharya}}]{Pautrat:2005ku}%
  \BibitemOpen
  \bibfield  {author} {\bibinfo {author} {\bibfnamefont {A.}~\bibnamefont
  {Pautrat}}, \bibinfo {author} {\bibfnamefont {J.}~\bibnamefont {Scola}},
  \bibinfo {author} {\bibfnamefont {C.}~\bibnamefont {Simon}}, \bibinfo
  {author} {\bibfnamefont {A.}~\bibnamefont {Br{\^u}let}}, \bibinfo {author}
  {\bibfnamefont {C.}~\bibnamefont {Goupil}}, \bibinfo {author} {\bibfnamefont
  {M.~J.}\ \bibnamefont {Higgins}}, \ and\ \bibinfo {author} {\bibfnamefont
  {S.}~\bibnamefont {Bhattacharya}},\ }\href@noop {} {\bibfield  {journal}
  {\bibinfo  {journal} {Phys. Rev. B}\ }\textbf {\bibinfo {volume} {71}},\
  \bibinfo {pages} {064517} (\bibinfo {year} {2005})}\BibitemShut {NoStop}%
\bibitem [{\citenamefont {Li}\ \emph {et~al.}(2006)\citenamefont {Li},
  \citenamefont {Andrei}, \citenamefont {Xiao}, \citenamefont {Shuk},\ and\
  \citenamefont {Greenblatt}}]{Li:2006cg}%
  \BibitemOpen
  \bibfield  {author} {\bibinfo {author} {\bibfnamefont {G.}~\bibnamefont
  {Li}}, \bibinfo {author} {\bibfnamefont {E.~Y.}\ \bibnamefont {Andrei}},
  \bibinfo {author} {\bibfnamefont {Z.~L.}\ \bibnamefont {Xiao}}, \bibinfo
  {author} {\bibfnamefont {P.}~\bibnamefont {Shuk}}, \ and\ \bibinfo {author}
  {\bibfnamefont {M.}~\bibnamefont {Greenblatt}},\ }\href@noop {} {\bibfield
  {journal} {\bibinfo  {journal} {Phys. Rev. Lett.}\ }\textbf {\bibinfo
  {volume} {96}},\ \bibinfo {pages} {017009} (\bibinfo {year}
  {2006})}\BibitemShut {NoStop}%
\bibitem [{\citenamefont {Levett}\ \emph {et~al.}(2002)\citenamefont {Levett},
  \citenamefont {Dewhurst},\ and\ \citenamefont {Paul}}]{Levett:2002ba}%
  \BibitemOpen
  \bibfield  {author} {\bibinfo {author} {\bibfnamefont {S.~J.}\ \bibnamefont
  {Levett}}, \bibinfo {author} {\bibfnamefont {C.~D.}\ \bibnamefont
  {Dewhurst}}, \ and\ \bibinfo {author} {\bibfnamefont {D.~M.}\ \bibnamefont
  {Paul}},\ }\href@noop {} {\bibfield  {journal} {\bibinfo  {journal} {Phys.
  Rev. B}\ }\textbf {\bibinfo {volume} {66}},\ \bibinfo {pages} {014515}
  (\bibinfo {year} {2002})}\BibitemShut {NoStop}%
\bibitem [{\citenamefont {Louden}\ \emph
  {et~al.}(2019{\natexlab{a}})\citenamefont {Louden}, \citenamefont
  {Rastovski}, \citenamefont {Kuhn}, \citenamefont {Leishman}, \citenamefont
  {DeBeer-Schmitt}, \citenamefont {Dewhurst}, \citenamefont {Zhigadlo},\ and\
  \citenamefont {Eskildsen}}]{Louden:2019bq}%
  \BibitemOpen
  \bibfield  {author} {\bibinfo {author} {\bibfnamefont {E.~R.}\ \bibnamefont
  {Louden}}, \bibinfo {author} {\bibfnamefont {C.}~\bibnamefont {Rastovski}},
  \bibinfo {author} {\bibfnamefont {S.~J.}\ \bibnamefont {Kuhn}}, \bibinfo
  {author} {\bibfnamefont {A.~W.~D.}\ \bibnamefont {Leishman}}, \bibinfo
  {author} {\bibfnamefont {L.}~\bibnamefont {DeBeer-Schmitt}}, \bibinfo
  {author} {\bibfnamefont {C.~D.}\ \bibnamefont {Dewhurst}}, \bibinfo {author}
  {\bibfnamefont {N.~D.}\ \bibnamefont {Zhigadlo}}, \ and\ \bibinfo {author}
  {\bibfnamefont {M.~R.}\ \bibnamefont {Eskildsen}},\ }\href@noop {} {\bibfield
   {journal} {\bibinfo  {journal} {Phys. Rev. B}\ }\textbf {\bibinfo {volume}
  {99}},\ \bibinfo {pages} {060502(R)} (\bibinfo {year}
  {2019}{\natexlab{a}})}\BibitemShut {NoStop}%
\bibitem [{\citenamefont {Louden}\ \emph
  {et~al.}(2019{\natexlab{b}})\citenamefont {Louden}, \citenamefont
  {Rastovski}, \citenamefont {DeBeer-Schmitt}, \citenamefont {Dewhurst},
  \citenamefont {Zhigadlo},\ and\ \citenamefont {Eskildsen}}]{Louden:2019io}%
  \BibitemOpen
  \bibfield  {author} {\bibinfo {author} {\bibfnamefont {E.~R.}\ \bibnamefont
  {Louden}}, \bibinfo {author} {\bibfnamefont {C.}~\bibnamefont {Rastovski}},
  \bibinfo {author} {\bibfnamefont {L.}~\bibnamefont {DeBeer-Schmitt}},
  \bibinfo {author} {\bibfnamefont {C.~D.}\ \bibnamefont {Dewhurst}}, \bibinfo
  {author} {\bibfnamefont {N.~D.}\ \bibnamefont {Zhigadlo}}, \ and\ \bibinfo
  {author} {\bibfnamefont {M.~R.}\ \bibnamefont {Eskildsen}},\ }\href@noop {}
  {\bibfield  {journal} {\bibinfo  {journal} {Phys. Rev. B}\ }\textbf {\bibinfo
  {volume} {99}},\ \bibinfo {pages} {144515} (\bibinfo {year}
  {2019}{\natexlab{b}})}\BibitemShut {NoStop}%
\bibitem [{\citenamefont {Eskildsen}\ \emph {et~al.}(2009)\citenamefont
  {Eskildsen}, \citenamefont {Vinnikov}, \citenamefont {Blasius}, \citenamefont
  {Veshchunov}, \citenamefont {Artemova}, \citenamefont {Densmore},
  \citenamefont {Dewhurst}, \citenamefont {Ni}, \citenamefont {Kreyssig},
  \citenamefont {Bud{\textquoteright}ko}, \citenamefont {Canfield},\ and\
  \citenamefont {Goldman}}]{Eskildsen:2009cx}%
  \BibitemOpen
  \bibfield  {author} {\bibinfo {author} {\bibfnamefont {M.~R.}\ \bibnamefont
  {Eskildsen}}, \bibinfo {author} {\bibfnamefont {L.~Y.}\ \bibnamefont
  {Vinnikov}}, \bibinfo {author} {\bibfnamefont {T.~D.}\ \bibnamefont
  {Blasius}}, \bibinfo {author} {\bibfnamefont {I.~S.}\ \bibnamefont
  {Veshchunov}}, \bibinfo {author} {\bibfnamefont {T.~M.}\ \bibnamefont
  {Artemova}}, \bibinfo {author} {\bibfnamefont {J.~M.}\ \bibnamefont
  {Densmore}}, \bibinfo {author} {\bibfnamefont {C.~D.}\ \bibnamefont
  {Dewhurst}}, \bibinfo {author} {\bibfnamefont {N.}~\bibnamefont {Ni}},
  \bibinfo {author} {\bibfnamefont {A.}~\bibnamefont {Kreyssig}}, \bibinfo
  {author} {\bibfnamefont {S.~L.}\ \bibnamefont {Bud{\textquoteright}ko}},
  \bibinfo {author} {\bibfnamefont {P.~C.}\ \bibnamefont {Canfield}}, \ and\
  \bibinfo {author} {\bibfnamefont {A.~I.}\ \bibnamefont {Goldman}},\
  }\href@noop {} {\bibfield  {journal} {\bibinfo  {journal} {Phys. Rev. B}\
  }\textbf {\bibinfo {volume} {79}},\ \bibinfo {pages} {100501(R)} (\bibinfo
  {year} {2009})}\BibitemShut {NoStop}%
\bibitem [{\citenamefont {Inosov}\ \emph
  {et~al.}(2010{\natexlab{a}})\citenamefont {Inosov}, \citenamefont {Shapoval},
  \citenamefont {Neu}, \citenamefont {Wolff}, \citenamefont {White},
  \citenamefont {Haindl}, \citenamefont {Park}, \citenamefont {Sun},
  \citenamefont {Lin}, \citenamefont {Forgan}, \citenamefont {Viazovska},
  \citenamefont {Kim}, \citenamefont {Laver}, \citenamefont {Nenkov},
  \citenamefont {Khvostikova}, \citenamefont {Kuhnemann},\ and\ \citenamefont
  {Hinkov}}]{Inosov:2010aa}%
  \BibitemOpen
  \bibfield  {author} {\bibinfo {author} {\bibfnamefont {D.~S.}\ \bibnamefont
  {Inosov}}, \bibinfo {author} {\bibfnamefont {T.}~\bibnamefont {Shapoval}},
  \bibinfo {author} {\bibfnamefont {V.}~\bibnamefont {Neu}}, \bibinfo {author}
  {\bibfnamefont {U.}~\bibnamefont {Wolff}}, \bibinfo {author} {\bibfnamefont
  {J.~S.}\ \bibnamefont {White}}, \bibinfo {author} {\bibfnamefont
  {S.}~\bibnamefont {Haindl}}, \bibinfo {author} {\bibfnamefont {J.~T.}\
  \bibnamefont {Park}}, \bibinfo {author} {\bibfnamefont {D.~L.}\ \bibnamefont
  {Sun}}, \bibinfo {author} {\bibfnamefont {C.~T.}\ \bibnamefont {Lin}},
  \bibinfo {author} {\bibfnamefont {E.~M.}\ \bibnamefont {Forgan}}, \bibinfo
  {author} {\bibfnamefont {M.~S.}\ \bibnamefont {Viazovska}}, \bibinfo {author}
  {\bibfnamefont {J.~H.}\ \bibnamefont {Kim}}, \bibinfo {author} {\bibfnamefont
  {M.}~\bibnamefont {Laver}}, \bibinfo {author} {\bibfnamefont
  {K.}~\bibnamefont {Nenkov}}, \bibinfo {author} {\bibfnamefont
  {O.}~\bibnamefont {Khvostikova}}, \bibinfo {author} {\bibfnamefont
  {S.}~\bibnamefont {Kuhnemann}}, \ and\ \bibinfo {author} {\bibfnamefont
  {V.}~\bibnamefont {Hinkov}},\ }\href@noop {} {\bibfield  {journal} {\bibinfo
  {journal} {Phys. Rev. B}\ }\textbf {\bibinfo {volume} {81}},\ \bibinfo
  {pages} {014513} (\bibinfo {year} {2010}{\natexlab{a}})}\BibitemShut
  {NoStop}%
\bibitem [{\citenamefont {Inosov}\ \emph
  {et~al.}(2010{\natexlab{b}})\citenamefont {Inosov}, \citenamefont {White},
  \citenamefont {Evtushinsky}, \citenamefont {Morozov}, \citenamefont
  {Cameron}, \citenamefont {Stockert}, \citenamefont {Zabolotnyy},
  \citenamefont {Kim}, \citenamefont {Kordyuk}, \citenamefont {Borisenko},
  \citenamefont {Forgan}, \citenamefont {Klingeler}, \citenamefont {Park},
  \citenamefont {Wurmehl}, \citenamefont {Vasiliev}, \citenamefont {Behr},
  \citenamefont {Dewhurst},\ and\ \citenamefont {Hinkov}}]{Inosov:2010bb}%
  \BibitemOpen
  \bibfield  {author} {\bibinfo {author} {\bibfnamefont {D.~S.}\ \bibnamefont
  {Inosov}}, \bibinfo {author} {\bibfnamefont {J.~S.}\ \bibnamefont {White}},
  \bibinfo {author} {\bibfnamefont {D.~V.}\ \bibnamefont {Evtushinsky}},
  \bibinfo {author} {\bibfnamefont {I.~V.}\ \bibnamefont {Morozov}}, \bibinfo
  {author} {\bibfnamefont {A.}~\bibnamefont {Cameron}}, \bibinfo {author}
  {\bibfnamefont {U.}~\bibnamefont {Stockert}}, \bibinfo {author}
  {\bibfnamefont {V.~B.}\ \bibnamefont {Zabolotnyy}}, \bibinfo {author}
  {\bibfnamefont {T.~K.}\ \bibnamefont {Kim}}, \bibinfo {author} {\bibfnamefont
  {A.~A.}\ \bibnamefont {Kordyuk}}, \bibinfo {author} {\bibfnamefont {S.~V.}\
  \bibnamefont {Borisenko}}, \bibinfo {author} {\bibfnamefont {E.~M.}\
  \bibnamefont {Forgan}}, \bibinfo {author} {\bibfnamefont {R.}~\bibnamefont
  {Klingeler}}, \bibinfo {author} {\bibfnamefont {J.~T.}\ \bibnamefont {Park}},
  \bibinfo {author} {\bibfnamefont {S.}~\bibnamefont {Wurmehl}}, \bibinfo
  {author} {\bibfnamefont {A.~N.}\ \bibnamefont {Vasiliev}}, \bibinfo {author}
  {\bibfnamefont {G.}~\bibnamefont {Behr}}, \bibinfo {author} {\bibfnamefont
  {C.~D.}\ \bibnamefont {Dewhurst}}, \ and\ \bibinfo {author} {\bibfnamefont
  {V.}~\bibnamefont {Hinkov}},\ }\href@noop {} {\bibfield  {journal} {\bibinfo
  {journal} {Phys. Rev. Lett.}\ }\textbf {\bibinfo {volume} {104}},\ \bibinfo
  {pages} {187001} (\bibinfo {year} {2010}{\natexlab{b}})}\BibitemShut
  {NoStop}%
\bibitem [{\citenamefont {M\"uhlbauer}\ \emph {et~al.}(2019)\citenamefont
  {M\"uhlbauer}, \citenamefont {Honecker}, \citenamefont {P\'erigo},
  \citenamefont {Bergner}, \citenamefont {Disch}, \citenamefont {Heinemann},
  \citenamefont {Erokhin}, \citenamefont {Berkov}, \citenamefont {Leighton},
  \citenamefont {Eskildsen},\ and\ \citenamefont {Michels}}]{Muhlbauer:2019jt}%
  \BibitemOpen
  \bibfield  {author} {\bibinfo {author} {\bibfnamefont {S.}~\bibnamefont
  {M\"uhlbauer}}, \bibinfo {author} {\bibfnamefont {D.}~\bibnamefont
  {Honecker}}, \bibinfo {author} {\bibfnamefont {E.~A.}\ \bibnamefont
  {P\'erigo}}, \bibinfo {author} {\bibfnamefont {F.}~\bibnamefont {Bergner}},
  \bibinfo {author} {\bibfnamefont {S.}~\bibnamefont {Disch}}, \bibinfo
  {author} {\bibfnamefont {A.}~\bibnamefont {Heinemann}}, \bibinfo {author}
  {\bibfnamefont {S.}~\bibnamefont {Erokhin}}, \bibinfo {author} {\bibfnamefont
  {D.}~\bibnamefont {Berkov}}, \bibinfo {author} {\bibfnamefont
  {C.}~\bibnamefont {Leighton}}, \bibinfo {author} {\bibfnamefont {M.~R.}\
  \bibnamefont {Eskildsen}}, \ and\ \bibinfo {author} {\bibfnamefont
  {A.}~\bibnamefont {Michels}},\ }\href@noop {} {\bibfield  {journal} {\bibinfo
   {journal} {Rev. Mod. Phys}\ }\textbf {\bibinfo {volume} {91}},\ \bibinfo
  {pages} {015004} (\bibinfo {year} {2019})}\BibitemShut {NoStop}%
\bibitem [{\citenamefont {Heller}\ \emph {et~al.}(2018)\citenamefont {Heller},
  \citenamefont {Cuneo}, \citenamefont {Debeer-Schmitt}, \citenamefont {Do},
  \citenamefont {He}, \citenamefont {Heroux}, \citenamefont {Littrell},
  \citenamefont {Pingali}, \citenamefont {Qian}, \citenamefont {Stanley},
  \citenamefont {Urban}, \citenamefont {Wu},\ and\ \citenamefont
  {Bras}}]{Heller:2018}%
  \BibitemOpen
  \bibfield  {author} {\bibinfo {author} {\bibfnamefont {W.~T.}\ \bibnamefont
  {Heller}}, \bibinfo {author} {\bibfnamefont {M.}~\bibnamefont {Cuneo}},
  \bibinfo {author} {\bibfnamefont {L.}~\bibnamefont {Debeer-Schmitt}},
  \bibinfo {author} {\bibfnamefont {C.}~\bibnamefont {Do}}, \bibinfo {author}
  {\bibfnamefont {L.}~\bibnamefont {He}}, \bibinfo {author} {\bibfnamefont
  {L.}~\bibnamefont {Heroux}}, \bibinfo {author} {\bibfnamefont
  {K.}~\bibnamefont {Littrell}}, \bibinfo {author} {\bibfnamefont {S.~V.}\
  \bibnamefont {Pingali}}, \bibinfo {author} {\bibfnamefont {S.}~\bibnamefont
  {Qian}}, \bibinfo {author} {\bibfnamefont {C.}~\bibnamefont {Stanley}},
  \bibinfo {author} {\bibfnamefont {V.~S.}\ \bibnamefont {Urban}}, \bibinfo
  {author} {\bibfnamefont {B.}~\bibnamefont {Wu}}, \ and\ \bibinfo {author}
  {\bibfnamefont {W.}~\bibnamefont {Bras}},\ }\href@noop {} {\bibfield
  {journal} {\bibinfo  {journal} {J. Appl. Crystallogr.}\ }\textbf {\bibinfo
  {volume} {51}},\ \bibinfo {pages} {242} (\bibinfo {year} {2018})}\BibitemShut
  {NoStop}%
\bibitem [{\citenamefont {Dewhurst}(2014)}]{Dewhurst:2014wo}%
  \BibitemOpen
  \bibfield  {author} {\bibinfo {author} {\bibfnamefont {C.~D.}\ \bibnamefont
  {Dewhurst}},\ }\href@noop {} {\bibfield  {journal} {\bibinfo  {journal} {J.
  Appl. Cryst.}\ }\textbf {\bibinfo {volume} {47}},\ \bibinfo {pages} {1180}
  (\bibinfo {year} {2014})}\BibitemShut {NoStop}%
\bibitem [{\citenamefont {Eskildsen}\ \emph {et~al.}(2016)\citenamefont
  {Eskildsen}, \citenamefont {Avers}, \citenamefont {Dewhurst}, \citenamefont
  {Gannon}, \citenamefont {Halperin}, \citenamefont {Kuhn},\ and\ \citenamefont
  {White}}]{5-42-402}%
  \BibitemOpen
  \bibfield  {author} {\bibinfo {author} {\bibfnamefont {M.~R.}\ \bibnamefont
  {Eskildsen}}, \bibinfo {author} {\bibfnamefont {K.}~\bibnamefont {Avers}},
  \bibinfo {author} {\bibfnamefont {C.}~\bibnamefont {Dewhurst}}, \bibinfo
  {author} {\bibfnamefont {W.}~\bibnamefont {Gannon}}, \bibinfo {author}
  {\bibfnamefont {W.~P.}\ \bibnamefont {Halperin}}, \bibinfo {author}
  {\bibfnamefont {S.}~\bibnamefont {Kuhn}}, \ and\ \bibinfo {author}
  {\bibfnamefont {J.}~\bibnamefont {White}},\ }\href@noop {} {\enquote
  {\bibinfo {title} {{Chiral Effects on the Vortex Lattice in UPt$_3$}},}\
  }\bibinfo {howpublished} {Institut Laue-Langevin (ILL)
  doi:10.5291/ILL-DATA.5-42-402} (\bibinfo {year} {2016})\BibitemShut {NoStop}%
\bibitem [{\citenamefont {Eskildsen}\ \emph {et~al.}(2018)\citenamefont
  {Eskildsen}, \citenamefont {Avers}, \citenamefont {Dewhurst}, \citenamefont
  {Honecker}, \citenamefont {Leishman},\ and\ \citenamefont
  {White}}]{5-42-467}%
  \BibitemOpen
  \bibfield  {author} {\bibinfo {author} {\bibfnamefont {M.~R.}\ \bibnamefont
  {Eskildsen}}, \bibinfo {author} {\bibfnamefont {K.}~\bibnamefont {Avers}},
  \bibinfo {author} {\bibfnamefont {C.}~\bibnamefont {Dewhurst}}, \bibinfo
  {author} {\bibfnamefont {D.}~\bibnamefont {Honecker}}, \bibinfo {author}
  {\bibfnamefont {A.}~\bibnamefont {Leishman}}, \ and\ \bibinfo {author}
  {\bibfnamefont {J.}~\bibnamefont {White}},\ }\href@noop {} {\enquote
  {\bibinfo {title} {{Chiral Effects on the Vortex Lattice in UPt$_3$}},}\
  }\bibinfo {howpublished} {Institut Laue-Langevin (ILL)
  doi:10.5291/ILL-DATA.5-42-467} (\bibinfo {year} {2018})\BibitemShut {NoStop}%
\bibitem [{\citenamefont {Gannon}\ \emph {et~al.}(2015)\citenamefont {Gannon},
  \citenamefont {Halperin}, \citenamefont {Rastovski}, \citenamefont
  {Schlesinger}, \citenamefont {Hlevyack}, \citenamefont {Eskildsen},
  \citenamefont {Vorontsov}, \citenamefont {Gavilano}, \citenamefont {Gasser},\
  and\ \citenamefont {Nagy}}]{Gannon:2015ct}%
  \BibitemOpen
  \bibfield  {author} {\bibinfo {author} {\bibfnamefont {W.~J.}\ \bibnamefont
  {Gannon}}, \bibinfo {author} {\bibfnamefont {W.~P.}\ \bibnamefont
  {Halperin}}, \bibinfo {author} {\bibfnamefont {C.}~\bibnamefont {Rastovski}},
  \bibinfo {author} {\bibfnamefont {K.~J.}\ \bibnamefont {Schlesinger}},
  \bibinfo {author} {\bibfnamefont {J.}~\bibnamefont {Hlevyack}}, \bibinfo
  {author} {\bibfnamefont {M.~R.}\ \bibnamefont {Eskildsen}}, \bibinfo {author}
  {\bibfnamefont {A.~B.}\ \bibnamefont {Vorontsov}}, \bibinfo {author}
  {\bibfnamefont {J.}~\bibnamefont {Gavilano}}, \bibinfo {author}
  {\bibfnamefont {U.}~\bibnamefont {Gasser}}, \ and\ \bibinfo {author}
  {\bibfnamefont {G.}~\bibnamefont {Nagy}},\ }\href@noop {} {\bibfield
  {journal} {\bibinfo  {journal} {New J. Phys.}\ }\textbf {\bibinfo {volume}
  {17}},\ \bibinfo {pages} {023041} (\bibinfo {year} {2015})}\BibitemShut
  {NoStop}%
\bibitem [{\citenamefont {Avers}\ \emph {et~al.}(2020)\citenamefont {Avers},
  \citenamefont {Gannon}, \citenamefont {Kuhn}, \citenamefont {Halperin},
  \citenamefont {Sauls}, \citenamefont {DeBeer-Schmitt}, \citenamefont
  {Dewhurst}, \citenamefont {Gavilano}, \citenamefont {Nagy}, \citenamefont
  {Gasser},\ and\ \citenamefont {Eskildsen}}]{Avers:2020wx}%
  \BibitemOpen
  \bibfield  {author} {\bibinfo {author} {\bibfnamefont {K.~E.}\ \bibnamefont
  {Avers}}, \bibinfo {author} {\bibfnamefont {W.~J.}\ \bibnamefont {Gannon}},
  \bibinfo {author} {\bibfnamefont {S.~J.}\ \bibnamefont {Kuhn}}, \bibinfo
  {author} {\bibfnamefont {W.~P.}\ \bibnamefont {Halperin}}, \bibinfo {author}
  {\bibfnamefont {J.~A.}\ \bibnamefont {Sauls}}, \bibinfo {author}
  {\bibfnamefont {L.}~\bibnamefont {DeBeer-Schmitt}}, \bibinfo {author}
  {\bibfnamefont {C.~D.}\ \bibnamefont {Dewhurst}}, \bibinfo {author}
  {\bibfnamefont {J.}~\bibnamefont {Gavilano}}, \bibinfo {author}
  {\bibfnamefont {G.}~\bibnamefont {Nagy}}, \bibinfo {author} {\bibfnamefont
  {U.}~\bibnamefont {Gasser}}, \ and\ \bibinfo {author} {\bibfnamefont {M.~R.}\
  \bibnamefont {Eskildsen}},\ }\href@noop {} {\bibfield  {journal} {\bibinfo
  {journal} {Nat. Phys.}\ }\textbf {\bibinfo {volume} {16}},\ \bibinfo {pages}
  {531} (\bibinfo {year} {2020})}\BibitemShut {NoStop}%
\bibitem [{\citenamefont {Avers}\ \emph {et~al.}(2022)\citenamefont {Avers},
  \citenamefont {Gannon}, \citenamefont {Leishman}, \citenamefont
  {DeBeer-Schmitt}, \citenamefont {Halperin},\ and\ \citenamefont
  {Eskildsen}}]{Avers:2022gh}%
  \BibitemOpen
  \bibfield  {author} {\bibinfo {author} {\bibfnamefont {K.~E.}\ \bibnamefont
  {Avers}}, \bibinfo {author} {\bibfnamefont {W.~J.}\ \bibnamefont {Gannon}},
  \bibinfo {author} {\bibfnamefont {A.~W.~D.}\ \bibnamefont {Leishman}},
  \bibinfo {author} {\bibfnamefont {L.}~\bibnamefont {DeBeer-Schmitt}},
  \bibinfo {author} {\bibfnamefont {W.~P.}\ \bibnamefont {Halperin}}, \ and\
  \bibinfo {author} {\bibfnamefont {M.~R.}\ \bibnamefont {Eskildsen}},\
  }\href@noop {} {\bibfield  {journal} {\bibinfo  {journal} {Front. Electron.
  Mater.}\ }\textbf {\bibinfo {volume} {2}},\ \bibinfo {pages} {878308}
  (\bibinfo {year} {2022})}\BibitemShut {NoStop}%
\bibitem [{\citenamefont {Cubitt}\ \emph {et~al.}(1993)\citenamefont {Cubitt},
  \citenamefont {Forgan}, \citenamefont {Yang}, \citenamefont {Lee},
  \citenamefont {Paul}, \citenamefont {Mook}, \citenamefont {Yethiraj},
  \citenamefont {Kes}, \citenamefont {Li}, \citenamefont {Menovsky},
  \citenamefont {Tarnawski},\ and\ \citenamefont {Mortensen}}]{Cubitt:1993aa}%
  \BibitemOpen
  \bibfield  {author} {\bibinfo {author} {\bibfnamefont {R.}~\bibnamefont
  {Cubitt}}, \bibinfo {author} {\bibfnamefont {E.~M.}\ \bibnamefont {Forgan}},
  \bibinfo {author} {\bibfnamefont {G.}~\bibnamefont {Yang}}, \bibinfo {author}
  {\bibfnamefont {S.~L.}\ \bibnamefont {Lee}}, \bibinfo {author} {\bibfnamefont
  {D.~M.}\ \bibnamefont {Paul}}, \bibinfo {author} {\bibfnamefont {H.~A.}\
  \bibnamefont {Mook}}, \bibinfo {author} {\bibfnamefont {M.}~\bibnamefont
  {Yethiraj}}, \bibinfo {author} {\bibfnamefont {P.~H.}\ \bibnamefont {Kes}},
  \bibinfo {author} {\bibfnamefont {T.~W.}\ \bibnamefont {Li}}, \bibinfo
  {author} {\bibfnamefont {A.~A.}\ \bibnamefont {Menovsky}}, \bibinfo {author}
  {\bibfnamefont {Z.}~\bibnamefont {Tarnawski}}, \ and\ \bibinfo {author}
  {\bibfnamefont {K.}~\bibnamefont {Mortensen}},\ }\href@noop {} {\bibfield
  {journal} {\bibinfo  {journal} {Nature}\ }\textbf {\bibinfo {volume} {365}},\
  \bibinfo {pages} {407} (\bibinfo {year} {1993})}\BibitemShut {NoStop}%
\bibitem [{\citenamefont {Klein}\ \emph {et~al.}(2001)\citenamefont {Klein},
  \citenamefont {Joumard}, \citenamefont {Blanchard}, \citenamefont {Marcus},
  \citenamefont {Cubitt}, \citenamefont {Giamarchi},\ and\ \citenamefont
  {Le~Doussal}}]{Klein:2001aa}%
  \BibitemOpen
  \bibfield  {author} {\bibinfo {author} {\bibfnamefont {T.}~\bibnamefont
  {Klein}}, \bibinfo {author} {\bibfnamefont {I.}~\bibnamefont {Joumard}},
  \bibinfo {author} {\bibfnamefont {S.}~\bibnamefont {Blanchard}}, \bibinfo
  {author} {\bibfnamefont {J.}~\bibnamefont {Marcus}}, \bibinfo {author}
  {\bibfnamefont {R.}~\bibnamefont {Cubitt}}, \bibinfo {author} {\bibfnamefont
  {T.}~\bibnamefont {Giamarchi}}, \ and\ \bibinfo {author} {\bibfnamefont
  {P.}~\bibnamefont {Le~Doussal}},\ }\href@noop {} {\bibfield  {journal}
  {\bibinfo  {journal} {Nature}\ }\textbf {\bibinfo {volume} {413}},\ \bibinfo
  {pages} {404} (\bibinfo {year} {2001})}\BibitemShut {NoStop}%
\bibitem [{\citenamefont {Divakar}\ \emph {et~al.}(2004)\citenamefont
  {Divakar}, \citenamefont {Drew}, \citenamefont {Lee}, \citenamefont
  {Gilardi}, \citenamefont {Mesot}, \citenamefont {Ogrin}, \citenamefont
  {Charalambous}, \citenamefont {Forgan}, \citenamefont {Menon}, \citenamefont
  {Momono}, \citenamefont {Oda}, \citenamefont {Dewhurst},\ and\ \citenamefont
  {Baines}}]{Divakar:2004by}%
  \BibitemOpen
  \bibfield  {author} {\bibinfo {author} {\bibfnamefont {U.}~\bibnamefont
  {Divakar}}, \bibinfo {author} {\bibfnamefont {A.~J.}\ \bibnamefont {Drew}},
  \bibinfo {author} {\bibfnamefont {S.~L.}\ \bibnamefont {Lee}}, \bibinfo
  {author} {\bibfnamefont {R.}~\bibnamefont {Gilardi}}, \bibinfo {author}
  {\bibfnamefont {J.}~\bibnamefont {Mesot}}, \bibinfo {author} {\bibfnamefont
  {F.~Y.}\ \bibnamefont {Ogrin}}, \bibinfo {author} {\bibfnamefont
  {D.}~\bibnamefont {Charalambous}}, \bibinfo {author} {\bibfnamefont {E.~M.}\
  \bibnamefont {Forgan}}, \bibinfo {author} {\bibfnamefont {G.~I.}\
  \bibnamefont {Menon}}, \bibinfo {author} {\bibfnamefont {N.}~\bibnamefont
  {Momono}}, \bibinfo {author} {\bibfnamefont {M.}~\bibnamefont {Oda}},
  \bibinfo {author} {\bibfnamefont {C.~D.}\ \bibnamefont {Dewhurst}}, \ and\
  \bibinfo {author} {\bibfnamefont {C.}~\bibnamefont {Baines}},\ }\href@noop {}
  {\bibfield  {journal} {\bibinfo  {journal} {Phys. Rev. Lett.}\ }\textbf
  {\bibinfo {volume} {92}},\ \bibinfo {pages} {237004} (\bibinfo {year}
  {2004})}\BibitemShut {NoStop}%
\bibitem [{\citenamefont {Laver}\ \emph {et~al.}(2008)\citenamefont {Laver},
  \citenamefont {Forgan}, \citenamefont {Abrahamsen}, \citenamefont {Bowell},
  \citenamefont {Geue},\ and\ \citenamefont {Cubitt}}]{Laver:2008aa}%
  \BibitemOpen
  \bibfield  {author} {\bibinfo {author} {\bibfnamefont {M.}~\bibnamefont
  {Laver}}, \bibinfo {author} {\bibfnamefont {E.~M.}\ \bibnamefont {Forgan}},
  \bibinfo {author} {\bibfnamefont {A.~B.}\ \bibnamefont {Abrahamsen}},
  \bibinfo {author} {\bibfnamefont {C.}~\bibnamefont {Bowell}}, \bibinfo
  {author} {\bibfnamefont {T.}~\bibnamefont {Geue}}, \ and\ \bibinfo {author}
  {\bibfnamefont {R.}~\bibnamefont {Cubitt}},\ }\href@noop {} {\bibfield
  {journal} {\bibinfo  {journal} {Phys. Rev. Lett.}\ }\textbf {\bibinfo
  {volume} {100}},\ \bibinfo {pages} {107001} (\bibinfo {year}
  {2008})}\BibitemShut {NoStop}%
\bibitem [{\citenamefont {Toft-Petersen}\ \emph {et~al.}(2018)\citenamefont
  {Toft-Petersen}, \citenamefont {Abrahamsen}, \citenamefont {Balog},
  \citenamefont {Porcar},\ and\ \citenamefont {Laver}}]{ToftPetersen:2018ch}%
  \BibitemOpen
  \bibfield  {author} {\bibinfo {author} {\bibfnamefont {R.}~\bibnamefont
  {Toft-Petersen}}, \bibinfo {author} {\bibfnamefont {A.~B.}\ \bibnamefont
  {Abrahamsen}}, \bibinfo {author} {\bibfnamefont {S.}~\bibnamefont {Balog}},
  \bibinfo {author} {\bibfnamefont {L.}~\bibnamefont {Porcar}}, \ and\ \bibinfo
  {author} {\bibfnamefont {M.}~\bibnamefont {Laver}},\ }\href@noop {}
  {\bibfield  {journal} {\bibinfo  {journal} {Nat. Commun.}\ }\textbf {\bibinfo
  {volume} {9}},\ \bibinfo {pages} {901} (\bibinfo {year} {2018})}\BibitemShut
  {NoStop}%
\bibitem [{\citenamefont {Pedersen}\ \emph {et~al.}(1990)\citenamefont
  {Pedersen}, \citenamefont {Posselt},\ and\ \citenamefont
  {Mortensen}}]{Pedersen:1990aa}%
  \BibitemOpen
  \bibfield  {author} {\bibinfo {author} {\bibfnamefont {J.~S.}\ \bibnamefont
  {Pedersen}}, \bibinfo {author} {\bibfnamefont {D.}~\bibnamefont {Posselt}}, \
  and\ \bibinfo {author} {\bibfnamefont {K.}~\bibnamefont {Mortensen}},\
  }\href@noop {} {\bibfield  {journal} {\bibinfo  {journal} {J. Appl.
  Crystallogr.}\ }\textbf {\bibinfo {volume} {23}},\ \bibinfo {pages} {321}
  (\bibinfo {year} {1990})}\BibitemShut {NoStop}%
\bibitem [{\citenamefont {Daniilidis}\ \emph {et~al.}(2007)\citenamefont
  {Daniilidis}, \citenamefont {Park}, \citenamefont {Dimitrov}, \citenamefont
  {Lynn},\ and\ \citenamefont {Ling}}]{Daniilidis:2007wa}%
  \BibitemOpen
  \bibfield  {author} {\bibinfo {author} {\bibfnamefont {N.~D.}\ \bibnamefont
  {Daniilidis}}, \bibinfo {author} {\bibfnamefont {S.~R.}\ \bibnamefont
  {Park}}, \bibinfo {author} {\bibfnamefont {I.~K.}\ \bibnamefont {Dimitrov}},
  \bibinfo {author} {\bibfnamefont {J.~W.}\ \bibnamefont {Lynn}}, \ and\
  \bibinfo {author} {\bibfnamefont {X.~S.}\ \bibnamefont {Ling}},\ }\href@noop
  {} {\bibfield  {journal} {\bibinfo  {journal} {Phys. Rev. Lett.}\ }\textbf
  {\bibinfo {volume} {99}},\ \bibinfo {pages} {147007} (\bibinfo {year}
  {2007})}\BibitemShut {NoStop}%
\bibitem [{\citenamefont {Gammel}\ \emph {et~al.}(1998)\citenamefont {Gammel},
  \citenamefont {Yaron}, \citenamefont {Ramirez}, \citenamefont {Bishop},
  \citenamefont {Chang}, \citenamefont {Ruel}, \citenamefont {Pfeiffer},
  \citenamefont {Bucher}, \citenamefont {D'Anna}, \citenamefont {Huse},
  \citenamefont {Mortensen}, \citenamefont {Eskildsen},\ and\ \citenamefont
  {Kes}}]{Gammel:1998wj}%
  \BibitemOpen
  \bibfield  {author} {\bibinfo {author} {\bibfnamefont {P.~L.}\ \bibnamefont
  {Gammel}}, \bibinfo {author} {\bibfnamefont {U.}~\bibnamefont {Yaron}},
  \bibinfo {author} {\bibfnamefont {A.~P.}\ \bibnamefont {Ramirez}}, \bibinfo
  {author} {\bibfnamefont {D.~J.}\ \bibnamefont {Bishop}}, \bibinfo {author}
  {\bibfnamefont {A.~M.}\ \bibnamefont {Chang}}, \bibinfo {author}
  {\bibfnamefont {R.}~\bibnamefont {Ruel}}, \bibinfo {author} {\bibfnamefont
  {L.~N.}\ \bibnamefont {Pfeiffer}}, \bibinfo {author} {\bibfnamefont
  {E.}~\bibnamefont {Bucher}}, \bibinfo {author} {\bibfnamefont
  {G.}~\bibnamefont {D'Anna}}, \bibinfo {author} {\bibfnamefont {D.~A.}\
  \bibnamefont {Huse}}, \bibinfo {author} {\bibfnamefont {K.}~\bibnamefont
  {Mortensen}}, \bibinfo {author} {\bibfnamefont {M.~R.}\ \bibnamefont
  {Eskildsen}}, \ and\ \bibinfo {author} {\bibfnamefont {P.~H.}\ \bibnamefont
  {Kes}},\ }\href@noop {} {\bibfield  {journal} {\bibinfo  {journal} {Phys.
  Rev. Lett.}\ }\textbf {\bibinfo {volume} {80}},\ \bibinfo {pages} {833}
  (\bibinfo {year} {1998})}\BibitemShut {NoStop}%
\bibitem [{\citenamefont {Joumard}\ \emph {et~al.}(1999)\citenamefont
  {Joumard}, \citenamefont {Marcus}, \citenamefont {Klein},\ and\ \citenamefont
  {Cubitt}}]{Joumard:1999aa}%
  \BibitemOpen
  \bibfield  {author} {\bibinfo {author} {\bibfnamefont {I.}~\bibnamefont
  {Joumard}}, \bibinfo {author} {\bibfnamefont {J.}~\bibnamefont {Marcus}},
  \bibinfo {author} {\bibfnamefont {T.}~\bibnamefont {Klein}}, \ and\ \bibinfo
  {author} {\bibfnamefont {R.}~\bibnamefont {Cubitt}},\ }\href@noop {}
  {\bibfield  {journal} {\bibinfo  {journal} {Phys. Rev. Lett.}\ }\textbf
  {\bibinfo {volume} {82}},\ \bibinfo {pages} {4930} (\bibinfo {year}
  {1999})}\BibitemShut {NoStop}%
\bibitem [{\citenamefont {Yaron}\ \emph {et~al.}(1997)\citenamefont {Yaron},
  \citenamefont {Gammel}, \citenamefont {Boebinger}, \citenamefont {Aeppli},
  \citenamefont {Schiffer}, \citenamefont {Bucher}, \citenamefont {Bishop},
  \citenamefont {Broholm},\ and\ \citenamefont {Mortensen}}]{Yaron:1997wx}%
  \BibitemOpen
  \bibfield  {author} {\bibinfo {author} {\bibfnamefont {U.}~\bibnamefont
  {Yaron}}, \bibinfo {author} {\bibfnamefont {P.~L.}\ \bibnamefont {Gammel}},
  \bibinfo {author} {\bibfnamefont {G.~S.}\ \bibnamefont {Boebinger}}, \bibinfo
  {author} {\bibfnamefont {G.}~\bibnamefont {Aeppli}}, \bibinfo {author}
  {\bibfnamefont {P.}~\bibnamefont {Schiffer}}, \bibinfo {author}
  {\bibfnamefont {E.}~\bibnamefont {Bucher}}, \bibinfo {author} {\bibfnamefont
  {D.~J.}\ \bibnamefont {Bishop}}, \bibinfo {author} {\bibfnamefont
  {C.}~\bibnamefont {Broholm}}, \ and\ \bibinfo {author} {\bibfnamefont
  {K.}~\bibnamefont {Mortensen}},\ }\href@noop {} {\bibfield  {journal}
  {\bibinfo  {journal} {Phys. Rev. Lett.}\ }\textbf {\bibinfo {volume} {78}},\
  \bibinfo {pages} {3185} (\bibinfo {year} {1997})}\BibitemShut {NoStop}%
\bibitem [{\citenamefont {Joynt}\ and\ \citenamefont
  {Taillefer}(2002)}]{Joynt:2002wt}%
  \BibitemOpen
  \bibfield  {author} {\bibinfo {author} {\bibfnamefont {R.}~\bibnamefont
  {Joynt}}\ and\ \bibinfo {author} {\bibfnamefont {L.}~\bibnamefont
  {Taillefer}},\ }\href@noop {} {\bibfield  {journal} {\bibinfo  {journal}
  {Rev. Mod. Phys.}\ }\textbf {\bibinfo {volume} {74}},\ \bibinfo {pages} {235}
  (\bibinfo {year} {2002})}\BibitemShut {NoStop}%
\bibitem [{\citenamefont {Ramirez}\ \emph {et~al.}(1995)\citenamefont
  {Ramirez}, \citenamefont {St{\"u}cheli},\ and\ \citenamefont
  {Bucher}}]{Ramirez:1995fh}%
  \BibitemOpen
  \bibfield  {author} {\bibinfo {author} {\bibfnamefont {A.~P.}\ \bibnamefont
  {Ramirez}}, \bibinfo {author} {\bibfnamefont {N.}~\bibnamefont
  {St{\"u}cheli}}, \ and\ \bibinfo {author} {\bibfnamefont {E.}~\bibnamefont
  {Bucher}},\ }\href@noop {} {\bibfield  {journal} {\bibinfo  {journal} {Phys.
  Rev. Lett.}\ }\textbf {\bibinfo {volume} {74}},\ \bibinfo {pages} {1218}
  (\bibinfo {year} {1995})}\BibitemShut {NoStop}%
\bibitem [{\citenamefont {Suderow}\ \emph {et~al.}(1997)\citenamefont
  {Suderow}, \citenamefont {Brison}, \citenamefont {Huxley},\ and\
  \citenamefont {Flouquet}}]{Suderow:1997ex}%
  \BibitemOpen
  \bibfield  {author} {\bibinfo {author} {\bibfnamefont {H.}~\bibnamefont
  {Suderow}}, \bibinfo {author} {\bibfnamefont {J.~P.}\ \bibnamefont {Brison}},
  \bibinfo {author} {\bibfnamefont {A.}~\bibnamefont {Huxley}}, \ and\ \bibinfo
  {author} {\bibfnamefont {J.}~\bibnamefont {Flouquet}},\ }\href@noop {}
  {\bibfield  {journal} {\bibinfo  {journal} {J. Low Temp. Phys.}\ }\textbf
  {\bibinfo {volume} {108}},\ \bibinfo {pages} {11} (\bibinfo {year}
  {1997})}\BibitemShut {NoStop}%
\bibitem [{\citenamefont {{Estimated using SRIM software package}}()}]{SRIM}%
  \BibitemOpen
  \bibfield  {author} {\bibinfo {author} {\bibnamefont {{Estimated using SRIM
  software package}}},\ }\href@noop {} {}\bibinfo {howpublished}
  {\url{http://www.srim.org}}\BibitemShut {NoStop}%
\bibitem [{\citenamefont {Hong}(1999)}]{Hong:1999ud}%
  \BibitemOpen
  \bibfield  {author} {\bibinfo {author} {\bibfnamefont {J.-I.}\ \bibnamefont
  {Hong}},\ }\emph {\bibinfo {title} {{Strucure-Property Relationships for a
  Heavy Fermion Superconductor UPt$_{3}$}}},\ \href@noop {} {Ph.D. thesis},\
  \bibinfo  {school} {Northwestern University} (\bibinfo {year}
  {1999})\BibitemShut {NoStop}%
\bibitem [{\citenamefont {Bean}\ \emph {et~al.}(1966)\citenamefont {Bean},
  \citenamefont {Fleischer}, \citenamefont {Swartz},\ and\ \citenamefont
  {Hart~Jr}}]{Bean:1966dv}%
  \BibitemOpen
  \bibfield  {author} {\bibinfo {author} {\bibfnamefont {C.~P.}\ \bibnamefont
  {Bean}}, \bibinfo {author} {\bibfnamefont {R.~L.}\ \bibnamefont {Fleischer}},
  \bibinfo {author} {\bibfnamefont {P.~S.}\ \bibnamefont {Swartz}}, \ and\
  \bibinfo {author} {\bibfnamefont {H.~R.}\ \bibnamefont {Hart~Jr}},\
  }\href@noop {} {\bibfield  {journal} {\bibinfo  {journal} {J. Appl. Phys.}\
  }\textbf {\bibinfo {volume} {37}},\ \bibinfo {pages} {2218} (\bibinfo {year}
  {1966})}\BibitemShut {NoStop}%
\bibitem [{\citenamefont {Fleischer}\ \emph {et~al.}(1989)\citenamefont
  {Fleischer}, \citenamefont {H~R~Hart}, \citenamefont {Lay},\ and\
  \citenamefont {Luborsky}}]{Fleischer:1989gd}%
  \BibitemOpen
  \bibfield  {author} {\bibinfo {author} {\bibfnamefont {R.~L.}\ \bibnamefont
  {Fleischer}}, \bibinfo {author} {\bibfnamefont {J.}~\bibnamefont {H~R~Hart}},
  \bibinfo {author} {\bibfnamefont {K.~W.}\ \bibnamefont {Lay}}, \ and\
  \bibinfo {author} {\bibfnamefont {F.~E.}\ \bibnamefont {Luborsky}},\
  }\href@noop {} {\bibfield  {journal} {\bibinfo  {journal} {Phys. Rev. B}\
  }\textbf {\bibinfo {volume} {40}},\ \bibinfo {pages} {2163} (\bibinfo {year}
  {1989})}\BibitemShut {NoStop}%
\bibitem [{\citenamefont {Shan}\ \emph {et~al.}(2003)\citenamefont {Shan},
  \citenamefont {Wen},\ and\ \citenamefont {Dou}}]{Shan:2003ds}%
  \BibitemOpen
  \bibfield  {author} {\bibinfo {author} {\bibfnamefont {L.}~\bibnamefont
  {Shan}}, \bibinfo {author} {\bibfnamefont {H.-H.}\ \bibnamefont {Wen}}, \
  and\ \bibinfo {author} {\bibfnamefont {S.~X.}\ \bibnamefont {Dou}},\
  }\href@noop {} {\bibfield  {journal} {\bibinfo  {journal} {Physica C}\
  }\textbf {\bibinfo {volume} {390}},\ \bibinfo {pages} {80} (\bibinfo {year}
  {2003})}\BibitemShut {NoStop}%
\bibitem [{\citenamefont {Strand}\ \emph {et~al.}(2009)\citenamefont {Strand},
  \citenamefont {Van~Harlingen}, \citenamefont {Kycia},\ and\ \citenamefont
  {Halperin}}]{Strand:2009eq}%
  \BibitemOpen
  \bibfield  {author} {\bibinfo {author} {\bibfnamefont {J.~D.}\ \bibnamefont
  {Strand}}, \bibinfo {author} {\bibfnamefont {D.~J.}\ \bibnamefont
  {Van~Harlingen}}, \bibinfo {author} {\bibfnamefont {J.~B.}\ \bibnamefont
  {Kycia}}, \ and\ \bibinfo {author} {\bibfnamefont {W.~P.}\ \bibnamefont
  {Halperin}},\ }\href@noop {} {\bibfield  {journal} {\bibinfo  {journal}
  {Phys. Rev. Lett.}\ }\textbf {\bibinfo {volume} {103}},\ \bibinfo {pages}
  {197002} (\bibinfo {year} {2009})}\BibitemShut {NoStop}%
\bibitem [{\citenamefont {Schemm}\ \emph {et~al.}(2014)\citenamefont {Schemm},
  \citenamefont {Gannon}, \citenamefont {Wishne}, \citenamefont {Halperin},\
  and\ \citenamefont {Kapitulnik}}]{Schemm:2014fv}%
  \BibitemOpen
  \bibfield  {author} {\bibinfo {author} {\bibfnamefont {E.~R.}\ \bibnamefont
  {Schemm}}, \bibinfo {author} {\bibfnamefont {W.~J.}\ \bibnamefont {Gannon}},
  \bibinfo {author} {\bibfnamefont {C.~M.}\ \bibnamefont {Wishne}}, \bibinfo
  {author} {\bibfnamefont {W.~P.}\ \bibnamefont {Halperin}}, \ and\ \bibinfo
  {author} {\bibfnamefont {A.}~\bibnamefont {Kapitulnik}},\ }\href@noop {}
  {\bibfield  {journal} {\bibinfo  {journal} {Science}\ }\textbf {\bibinfo
  {volume} {345}},\ \bibinfo {pages} {190} (\bibinfo {year}
  {2014})}\BibitemShut {NoStop}%
\end{thebibliography}%

\newpage
\hspace{1cm}

\newpage
\section*{Supplemental Material}

\subsection{{\UPt} single crystals}
Both ZR8 and ZR11 are long, rod-like crystals and were cut into two pieces which were co-aligned and fixed with silver epoxy (EPOTEK E4110) to a copper cold finger for the SANS measurements.
For the SANS experiments the respective sample assembly was mounted onto the mixing chamber of a dilution refrigerator and placed inside a superconducting magnet, oriented with the crystalline $\textbf{a}$ axis vertical and the $\textbf{c}$ axis horizontally along the magnetic field and the neutron beam.
For both samples the neutron beam was masked off to illuminate a $7 \times 11$~mm$^2$ area.

Sample properties of the two {\UPt} single crystals are listed in Table~\ref{Xtals}.
These were determined from resistive measurements performed on small samples cut from the main crystals.
Here, RRR is the residual resistivity ratio, $T_{\rm c}$ is the superconducting transition temperature and $\Delta T_{\rm c}$ is the width of the transition.
The greater RRR and smaller $\Delta T_{\rm c}$ for ZR11 indicates its higher quality compared with ZR8.
\begin{table}[h]
    \begin{tabular}{lcccc}
    \hline \hline
    Sample \hspace{0.1cm} & \hspace{0.1cm} mass~(g) \hspace{0.1cm} & \hspace{0.1cm} RRR \hspace{0.1cm} & \hspace{0.1cm} $T_{\rm c}$~(mK) \hspace{0.1cm} & \hspace{0.1cm} $\Delta T_{\rm c}$~(mK) \\ \hline
    ZR8  & 15 & $> 600$ & $560 \pm 2$ & 10 \\
    ZR11 &  9 & $> 900$ & $557 \pm 2$ &  5 \\ \hline \hline
    \end{tabular}
    \caption{\label{Xtals}
        Properties of the two {\UPt} single crystals used for the SANS experiments.}
\end{table}

\subsection{Neutron absorption}
The measured transmission is $T = 35\%$ and $T = 49\%$ for ZR8 and ZR11 respectively, and the rate of neutron absorption is determined from the flux of the un-diffracted beam during the VL measurements by $n_{\text{abs}} = (1-T)/T \times n_{\text{undif}}$.
The average sample thickness along the beam direction is obtained from $T = e^{-\langle t \rangle/\tau}$ with $\tau = (\lambda_n/0.18\;\mbox{nm}) \sum \sigma_{\text{abs}}/V_{\text{f.u.}}$, where $\lambda_n = 0.75$~nm is the neutron wavelength, and $V_{\text{f.u.}}$ and $\sum \sigma_{\text{abs}} = 0.003 \sigma_{^{235}\text{U}} + 0.997 \sigma_{^{238}\text{U}} + 3 \sigma_{\text{Pt}}$ is the volume and total absorption cross section for one formula unit.
Here, the $^{235}$U concentration was determined from the fissile heating as described below.
This yields $\langle t \rangle_{\text{ZR8}} = 5.0$~mm and $\langle t \rangle_{\text{ZR11}} = 3.4$~mm, which allows a determination of the absorbed neutrons per unit volume when combined with the illuminated sample area.

\subsection{Fission event rate}
A determination of the fission event rate is difficult to obtain from the sample transmission since the beam attenuation is dominated by non-fissile absorption by $^{238}$U and Pt.
Instead, this is estimated from the fissile heating of the ZR8 sample, where an incident flux of $1.1 \times 10^6$ neutrons/second cause a $\sim 1.3$~$\mu$W reduction of the heating power required to maintain the dilution refrigerator mixing chamber at a constant temperature.
This greatly exceeds the $0.2$~fW kinetic energy flux of the incident neutrons.
Assuming a net energy release of 200~MeV, the observed heating yields $4\times10^4$ fission events/second.
This corresponds to roughly 6\% of the incident neutrons, and a $^{235}$U concentration of $\sim 0.3$\% which is typical for depleted uranium.
Since the ZR8 and ZR11 crystals were grown from the same batch of uranium, they are expected to have the same concentration of $^{235}$U.
The difference in the disordering rate between the two samples in Fig.~4(b) is therefore not due to different fission rates.

\subsection{Thermalization and average sample temperature}
The neutron beam induced fission will cause local heating, and it is important to consider whether this gives rise to a gradual increase of the sample temperature on the same time scale as the SANS measurements and thereby contributing to the decreasing VL scattered intensity.
Here we note that the largest thermal resistance is likely between the $^3$He/$^4$He dilution refrigerator mixture and the mixing chamber itself, rather than between the sample and the copper cold finger to which it is attached.
The thermalization of the sample as a whole is therefore expected to occur on the same second time scale as the response of the heating power applied to the copper cold finger, when the neutron beam is turned on or off.
This is 2-3 orders of magnitude faster than the observed decay of the VL intensity.

\begin{figure*}
	\includegraphics{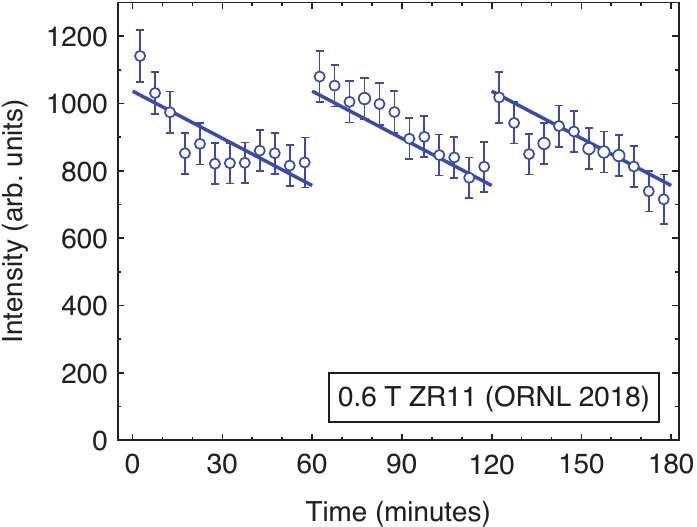}
	\caption{\label{DecayRepeat}
        Intensity of the VL Bragg peaks at 0.6~T measured in 5 minute time intervals.
        The neutron beam was continuously on during the measurements, and a damped field oscillation was applied at 0, 60, and 120 minutes to ``reset'' the VL.
        A 3 data point moving box-car average was used for each 60 minute sequence separately.
        The line shows a linear fit, repeated 3 times, to the three 60-minute sequences combined.
        Adding the intensities for the sequences yields $10,565 \pm 351$, $11,200 \pm 352$, and $10,415 \pm 350$ respectively, showing no systematic increase or decrease with time.
       }
\end{figure*}

To exclude the influence of a gradual increase of the average sample temperature we performed successive, time-dependent measurement of the decreasing VL intensity at the same field, shown in Fig.~\ref{DecayRepeat}.
From this it is clear that the VL intensity is restored by a damped field oscillation, and that the decay rate for successive measurements is the same within the experimental accuracy.
This definitively eliminates a gradual heating as the origin of the vanishing intensity with time.

\end{document}